\documentclass[useAMS,usenatbib]{mn2e}
\usepackage{graphicx}
\usepackage{booktabs}
\usepackage{subfigure}
\usepackage{ulem}
\usepackage{pdflscape}
\usepackage{amsmath,amssymb,latexsym}

\newcommand{\aap}{A\&A}
\newcommand{\aapr}{Astron. Astrophys. Rev.}
\newcommand{\aaps}{A\&A Suppl.}
\newcommand{\aca}{Acta Astron.}
\newcommand{\aj}{AJ}
\newcommand{\an}{AN}
\newcommand{\apj}{ApJ}
\newcommand{\apjl}{ApJL}
\newcommand{\apjs}{ApJ Suppl.}
\newcommand{\araa}{Ann. Rev. Astron. Astrophys.}
\newcommand{\aspc}{ASPC}
\newcommand{\astl}{Astron. Letters}
\newcommand{\ibvs}{Inf. Bull. on Var. Stars}
\newcommand{\jaavso}{J. of the American Assoc. of Var. Star Observers}
\newcommand{\Lsun}{\mathrm{L}_\odot}
\newcommand{\mnras}{MNRAS}
\newcommand{\Msun}{\mathrm{M}_\odot}
\newcommand{\pasa}{Publ. of the Astron. Soc. of Australia}
\newcommand{\pasp}{PASP}
\newcommand{\Rsun}{\mathrm{R}_\odot}
\newcommand{\sci}{Science}
\newcommand{\softL}{L\kern-0.8ex\raise0.1ex\hbox{'}\kern0.1ex}
\newcommand{\Teff}{T_\mathrm{eff}}
\newcommand{\tvspc}{\rule{0pt}{4ex}}

\title[T Tauri stars in the SuperWASP and NSVS surveys]{T Tauri stars in the SuperWASP and NSVS surveys
\thanks{Based on the data from SuperWASP and NSVS archives.}}

\author[{\softL}.\,Hamb\'alek et al.]
{{\softL}.\,Hamb\'alek$^1$\thanks{lhambalek@ta3.sk}, M.\,Va\v{n}ko$^1$,
E.\,Paunzen$^2$, B.\,Smalley$^3$\\
 $^1$Astronomical Institute, Slovak Academy of Sciences, 059 60 Tatransk\'a Lomnica, Slovakia\\
 $^2$Department of Theoretical Physics and Astrophysics, Masaryk University, Kotl\'a\v{r}ska 2, 61137 Brno, Czech Republic\\
 $^3$Astrophysics Group, Keele University, Keele ST5 5BG, UK\\}

\begin{document}

\date{Accepted 2018 February 28. Received 2018 January 19; in original form
2017 December 14}

\pagerange{\pageref{firstpage}--\pageref{lastpage}} \pubyear{2018}

\maketitle

\label{firstpage}

\begin{abstract}
We present a study of the long-term optical variability of young 
T~Tauri stars using previously unpublished data from the SuperWASP project.
Other publicly available photometry from NSVS and the NASA K2 mission were used
to check and supplement our results. 
Our sample includes twenty weak-lined T~Tauri stars in the Taurus-Auriga
star-forming region. We have performed a period search on the long time-series
photometry and derived the mean periods of stars in our sample.
We have found new periods for the stars V1334~Tau (HD~283782) and V1349~Tau
(HD~31281) without any period estimation in literature. The rotation period was
found for the primary star in the binary V773~Tau (HD~283447). Several earlier
results were updated.
For the star V410~Tau (HD~283518), we have compared the light curve changes
found in previous studies to the new measurements, and attributed the evolution
of spots to a $\sim$15-year cycle similar to the solar 11-year cycle. We have
also derived luminosities and effective temperatures for our targets, in order
to locate them in the Hertzsprung-Russell diagram and to calibrate the masses
and ages of the target stars.
\end{abstract}


\section{Introduction}
\label{intro}

The kinematic association of many T~Tauri stars with dark clouds where stars are
formed \citep{herbig77}, and the presence of Li\,I~$\lambda$6707 \AA~in
absorption \citep{bertout89} show that T~Tauri stars are young stellar objects.
They have low or intermediate masses (0.5$-$2.5 $\Msun$) and are defined as
pre-main-sequence (PMS) stars that are surrounded by a nebula and show emission
lines in their spectra \citep{joy45}. These objects typically have spectral
types ranging from G to M. PMS models show that their interiors are either fully
convective or possess outer convective envelopes, depending on the age and mass
of the star \citep{hussain12}. The properties of these PMS objects were reviewed
by, e.g. \citet{menber99} who emphasized two subgroups; the so-called
``classical'' T~Tauri stars (CTTS), still actively accreting from their
circumstellar disks and the ``weak-line'' T~Tauri stars (WTTS), no longer
surrounded by a circumstellar disk.

Depending on their spectral type, T~Tauri stars can have ages ranging from less
than one to tens of Myr. Comparing the position of these stars in the
HR diagram with theoretical evolutionary tracks (\citealp{dantona94, swenson94})
gives an upper-limit to the estimated age of the stars. However, these
evolutionary tracks do not take into account accretion and \citet{siess97}, show
that they underestimate ages by a factor of 2 to 3.

Ubiquitous amongst T~Tauri stars is their variability. As was already noticed in
the original definition by \citet{joy45}, the prototype T~Tauri exhibits strong
and mostly irregular variability on time-scales from hours to months and even
years. This is true for the broad-band photometry at all wavelengths (radio to
X-ray).
On top of a periodic signal due to star spots, modulated by stellar rotation,
the reason for this type of variability is stellar youth, apparent through
strong magnetic activity and variable accretion and/or extinction \citep{herbst94}.
Variations are not restricted to photometric observations. 
Emission lines are also changing in intensity and shape
(e.g., \citealp{johnsbasri95, lagogam98}), and polarimetric studies imply
variations in the degree and position angle of polarization as well
\citep{apenmun89}.

There have been many previous photometric surveys in the optical and infrared.
We refrain from presenting a comprehensive overview here, but mention several of
the surveys that have studied the photometric variability of our target stars.
\citet{bouvier97} observed in the optical range 58 WTTS which were detected in
the ROSAT All-Sky Survey (RASS). They were able to derive rotational periods for
18 of their stars, all but one being ascribed to rotational modulation by
stellar spots.
\citet{grankin08} presented a homogeneous set of photometric measurements for
WTTS extending for up to 20 years. Their data were collected within the framework
of the ROTOR (Research Of Traces Of Rotation) program, aimed at the study of the
photometric variability of PMS objects. The data
set contains rotational periods for 35 out of 48 stars. Further optical
photometry, including the behaviour on time-scales over more than several years, has
been studied in the works of \citet{gahm93}, \citet{grankin07}, \citet{percy10},
and \citet{ibryamov15}.

Recently, \citet{rigon17} presented a study, including long-term variability
of CTTS (mostly in the Taurus-Auriga region), based on data from the Wide Angle
Search for Planets (SuperWASP).
They found that the overwhelming majority of CTTS have a low-level variability
with $\sigma < 0.3$~mag dominated by time-scales of a few weeks, consistent
with rotational modulation. The presence of long-term variability correlates
with the spectral slope at 3--5~$\mu$m , which is an indicator of inner disc
geometry, and with the $U-B$ band slope, which is an accretion diagnostics.
This shows that the long-term variations in CTTS are predominantly driven by
processes in the inner disc and in the accretion zone.

The extensive simultaneous multiwavelength studies of \citealp{carpenter01}
find that the bulk of T~Tauri stars show photometric variability on the order
of 0.2~magnitudes in {\it JHKs} and $\sim$0.5~mag in their respective
near-infrared (NIR) colours. Further characterization in NIR was studied by, e.g.
\citet{eiroa02}, \citet{alves08}, \citet{rice12}. The photometric properties
have been extended to the mid-infrared (\citealp{flaherty12, rebull14}) and to
the far-infrared by \citep{billot12}.

The most advanced survey of T~Tauri stars to-date is the Coordinated Synoptic
Investigation's study of the NGC~2264 star-forming region (CSI~2264) \citep{cody14}.
The multi-wavelength observation campaign utilized 16 telescopes,
including the space-based observatories CoRoT and Spitzer. Their unprecedented
photometric precision $\leq1\%$ and a cadence down to several minutes sets a new
standard for this kind of surveys. \citealp{cody14, stauffer15, mcginnis15} show
that young-star variability can be caused not only by cold spots, but also by
circumstellar obscuration events, hot spots on the star and/or disk, accretion
bursts, and rapid structural changes in the inner disk.

The main goal of our study is to extend the knowledge of T~Tauri stars by
analysing several years of high-cadence monitoring of twenty WTTS using the data
from SuperWASP and NSVS surveys. Our particular emphasis is to search for
possible evolution in the long-term variability of brightness changes caused by
the presence of spots.

\section{Target selection}
\label{targets}

We have found that many bona fide T~Tauri stars (originally designated as members
of Tau-Aur region) in publications listed in Table~\ref{tab:basic} are without
reliable periods of photometric variability and/or physical parameters. We
excluded any known CTTS and stars with close visual companions from the list.
Furthermore, we selected only stars with $V<11$~mag, as this is the brightness
limit for spectroscopic follow-up observations at Star\'a Lesn\'a observatory
\citep{pribulla15}. We arrived to a sample of twenty WTTS.
All the targets have been observed by our own spectroscopic survey and the EW
of Li~$\lambda6707$ line was measured to confirm the evolutionary status of targets.
\citet{patterer93} found for HD~283518 (V410~Tau) EW(Li)=0.547$\pm$0.084~\AA.
All other stars in our sample had their EWs
measured in range of 0.165 -- 0.472~\AA.

\begin{table*}  
  \begin{center}
  \caption{The basic characteristic of the selected targets. Distances are
  calculated from parallaxes from the \textit{Gaia} DR2 \citep{lindegren18} catalogue.
  Distance of HD~286178 is from TGAS \citep{michalik15}. NSVS IDs with ``?''
  are close to another target and may have been misidentified.}
  \label{tab:basic}
  \begin{tabular}{lcccccccl}
  \hline
  Object    & SuperWASP ID              &  R.A.       & DEC        & $V_{\rm mag}$ & Sp. type & $P$ [d] & Dist. [pc]             & Refer.       \\
            & NSVS ID                   & (2000)      & (2000)     &               &          &         &                        &              \\
  \hline 
  HD~285281 & 1SWASPJ040031.06+193520.8 & 04 00 31.07 & 19 35 20.8 & 10.17         & K1       & 1.1683  & 135.3$^{+1.2}_{-1.2}$  & (1)(2)(6)    \\ 
            & 9414189?                  &             &            &               &          &         &                        &              \\
  \tvspc 
  BD+19~656 & 1SWASPJ040519.59+200925.5 & 04 05 19.61 & 20 09 25.2 & 10.12         & K1       & 2.86    & 108.5$^{+0.7}_{-0.7}$  & (2)(3)(6)    \\ 
            & 9417695?                  &             &            &               &          & (0.741) &                        &              \\
  \tvspc 
  HD~284135 & 1SWASPJ040540.58+224812.0 & 04 05 40.58 & 22 48 12.0 & 09.39         & G3V      & 0.8160  & --                     & (1)(6)       \\ 
            & 6750588                   &             &            &               &          &         &                        &              \\
  \tvspc 
  HD~284149 & 1SWASPJ040638.08+201811.1 & 04 06 38.80 & 20 18 11.2 & 09.63         & G0       & 1.0790  & 118.2$^{+0.7}_{-0.7}$  & (1)(2)(4)(6) \\ 
            & 9418722                   &             &            &               &          &         &                        &              \\
  \tvspc 
  HD~281691 & 1SWASPJ040909.74+290130.2 & 04 09 09.74 & 29 01 30.3 & 10.68         & G8III    & 2.74?   & 110.3$^{+0.5}_{-0.5}$  & (1)(6)       \\ 
            & 6754307                   &             &            &               &          &         &                        &              \\
  \tvspc 
  HD~284266 & 1SWASPJ041522.91+204417.0 & 04 15 22.92 & 20 44 16.9 & 10.51         & K0V      & 1.83    & 119.9$^{+1.0}_{-1.0}$  & (1)(2)(6)    \\ 
            & 9425183                   &             &            &               &          &         &                        &              \\
  \tvspc 
  HD~284503 & 1SWASPJ043049.18+211410.6 & 04 30 49.19 & 21 14 10.7 & 10.24         & G8       & 0.736   & 111.6$^{+0.7}_{-0.7}$  & (1)(6)       \\ 
            & 9436598                   &             &            &               &          &         &                        &              \\ 
  \tvspc 
  HD~284496 & 1SWASPJ043116.85+215025.2 & 04 31 16.86 & 21 50 25.3 & 10.80         & K0       & 2.71    & 125.8$^{+0.6}_{-0.6}$  & (1)(5)       \\ 
            & 9444066?                  &             &            &               &          &         &                        &              \\
  \tvspc 
  HD~285840 & 1SWASPJ043242.43+185510.2 & 04 32 42.43 & 18 55 10.2 & 10.85         & K1V      & 1.55    & 90.5$^{+0.3}_{-0.3}$   & (2)(5)(6)    \\ 
            & 9444857?                  &             &            &               &          &         &                        &              \\
  \tvspc 
  HD~285957 & 1SWASPJ043839.06+154613.6 & 04 38 39.07 & 15 46 13.6 & 10.86         & K1       & 3.07    & 139.2$^{+1.1}_{-1.1}$  & (2)(5)(6)    \\ 
            & 9449108                   &             &            &               &          &         &                        &              \\
  \tvspc 
  HD~283798 & 1SWASPJ044155.15+265849.4 & 04 41 55.16 & 26 58 49.4 & 09.55         & G7       & 0.6?    & 110.8$^{+0.6}_{-0.6}$  & (1)(2)(6)    \\ 
            & 6778011                   &             &            &               &          &         &                        &              \\
  \tvspc 
  HD~283782 & 1SWASPJ044454.45+271745.2 & 04 44 54.40 & 27 17 45.5 & 09.48         & K1       & ?       & 168.0$^{+6.8}_{-6.3}$  & (1)(2)(6)    \\ 
            & 6780256                   &             &            &               &          &         &                        &              \\
  \tvspc 
  HD~30171  & 1SWASPJ044551.29+155549.7 & 04 45 51.30 & 15 55 49.7 & 09.36         & G5       & 1.104   & 184.9$^{+3.9}_{-3.7}$  & (2)(5)(6)    \\ 
            & 9455701                   &             &            &               &          &         &                        &              \\
  \tvspc 
  HD~31281  & 1SWASPJ045509.62+182631.1 & 04 55 09.62 & 18 26 31.1 & 09.14         & G1       & ?       & 122.4$^{+0.6}_{-0.6}$  & (1)(2)(6)    \\ 
            & no data                   &             &            &               &          &         &                        &              \\
  \tvspc 
  HD~286179 & 1SWASPJ045700.64+151753.1 & 04 57 00.65 & 15 17 53.1 & 10.39         & G3       & 3.33    & 123.7$^{+1.0}_{-1.0}$  & (5)(6)       \\ 
            & 9466913                   &             &            &               &          &         &                        &              \\
  \tvspc 
  HD~286178 & 1SWASPJ045717.65+152509.4 & 04 57 17.66 & 15 25 09.5 & 10.54         & K1       &  2.39   & 74.3$^{+3.5}_{-3.2}$   & (2)(5)(6)    \\ 
            & 9467195                   &             &            &               &          &         &                        &              \\
  \tvspc 
  HD~283447 & 1SWASPJ041412.91+281212.3 & 04 14 12.92 & 28 12 12.3 & 10.68         & K3V      & 51      & 128.1$^{+2.3}_{-2.3}$  & (8)          \\ 
            & 6758228                   &             &            &               &          &         &                        &              \\ 
  \tvspc 
  HD~283572 & 1SWASPJ042158.84+281806.4 & 04 21 58.85 & 28 18 06.5 & 09.03         & G5       & 1.529   & 130.3$^{+0.9}_{-0.9}$  & (1)(2)       \\ 
            & no data                   &             &            &               &          &         &                        &              \\
  \tvspc 
  HD~285778 & 1SWASPJ042710.57+175042.6 & 04 27 10.57 & 17 50 42.6 & 10.22         & K1       & 2.734   & 120.1$^{+0.8}_{-0.8}$  & (7)          \\ 
            & 9440985                   &             &            &               &          &         &                        &              \\
  \tvspc 
  HD~283518 & 1SWASPJ041831.10+282716.0 & 04 18 31.12 & 28 27 16.1 & 10.75         & K3V      & 1.87    & 130.4$^{+0.9}_{-0.9}$  & (1)(9)(10)   \\ 
            & 6761377                   &             &            &               &          &         &                        &              \\ 
  \hline 
  \multicolumn{9}{l}{References: (1) \citet{grankin13}, (2) \citet{daemgen15}, (3) \citet{gran07}, (4) \citet{bonavita14},}\\
  \multicolumn{9}{l}{(5) \citet{bouvier97}, (6) \citet{wichmann00}, (7) \citet{grankin08}, (8) \citet{welty95}, (9) \citet{stelzer03},}\\
  \multicolumn{9}{l}{and (10) \citet{fernandez04}}\\
  \end{tabular}
  \end{center}
\end{table*}

\subsection{Evolution stage of target stars}
\label{hrd}

For the analysis of the evolutionary status of a star, the effective temperature
and luminosity is needed to locate the object in the Hertzsprung-Russell-diagram
(HRD). Then, the age and mass can be deduced from corresponding isochrones. 

\begin{table*}  
  \begin{center}
  \caption{The astrophysical parameters of our targets with available parallax measurements.} 
  \label{tab:stellar_parameters} 
  \begin{tabular}{lcccccc}
  \hline
  Object    & $A_{\rm V}$ & $\log \Teff$ & BC$_V$  & $\log L/\Lsun$ & $M$ [$\Msun$] & age [Myr] \\ 
  \hline   
  HD~285281 & 0.47(2)     & 3.699(22)    & -0.27   & +0.43(1)       & 1.4 -- 1.7    &  1 --  8  \\ 
  BD+19~656 & 0.27(4)     & 3.703(9)     & -0.26   & +0.07(2)       & 1.2 -- 1.3    &  7 -- 12  \\ 
  HD~284149 & 0.19(18)    & 3.775(11)    & -0.04   & +0.28(5)       & 1.0 -- 1.2    & 15 -- 25  \\ 
  HD~281691 & 0.19(11)    & 3.703(9)     & -0.26   & -0.01(4)       & 1.1 -- 1.3    &  8 -- 18  \\ 
  HD~284266 & 0.16(23)    & 3.724(20)    & -0.18   & -0.01(9)       & 1.0 -- 1.2    & 15 -- 30  \\ 
  HD~284503 & 0.19(5)     & 3.720(17)    & -0.19   & +0.05(2)       & 1.1 -- 1.3    & 10 -- 20  \\ 
  HD~284496 & 0.21(7)     & 3.716(13)    & -0.21   & +0.00(3)       & 1.1 -- 1.2    & 12 -- 20  \\ 
  HD~285840 & 0.17(25)    & 3.720(25)    & -0.19   & -0.33(10)      & 0.8 -- 1.0    & 20 -- 70  \\ 
  HD~285957 & 0.27(33)    & 3.695(13)    & -0.29   & +0.12(13)      & 1.2 -- 1.5    &  3 -- 13  \\ 
  HD~283798 & 0.00(1)     & 3.756(8)     & -0.09   & +0.19(1)       & 0.9 -- 1.3    & 17 -- 21  \\ 
  HD~283782 & 0.63(19)    & 3.716(29)    & -0.21   & +0.96(18)      & 1.8 -- 2.7    & $<3$      \\ 
  HD~30171  & 0.36(13)    & 3.736(16)    & -0.15   & +0.91(6)       & 2.1 -- 2.5    &  2 --  4  \\ 
  HD~31281  & 0.26(13)    & 3.763(7)     & -0.06   & +0.54(5)       & 1.4 -- 1.6    &  8 -- 12  \\ 
  HD~286179 & 0.44(29)    & 3.756(15)    & -0.09   & +0.19(12)      & 1.1 -- 1.4    & 10 -- 35  \\ 
  HD~283447 & 0.95(10)    & 3.690(13)    & -0.30   & +0.36(4)       & 1.4 -- 1.7    &  2 --  4  \\ 
  HD~283518 & 1.10(10)    & 3.643(44)    & -0.60   & +0.53(4)       & 0.5 -- 1.6    & $<2$      \\ 
  HD~283572 & 0.48(3)     & 3.740(24)    & -0.14   & +0.82(1)       & 1.8 -- 2.5    &  2 --  5  \\ 
  HD~285778 & 0.15(11)    & 3.720(12)    & -0.19   & +0.14(4)       & 1.2 -- 1.4    &  8 -- 15  \\ 
  \hline 
  \end{tabular}
  \end{center}
\end{table*}

\begin{figure*} 
  \centering
  \includegraphics[width=161mm,clip=0]{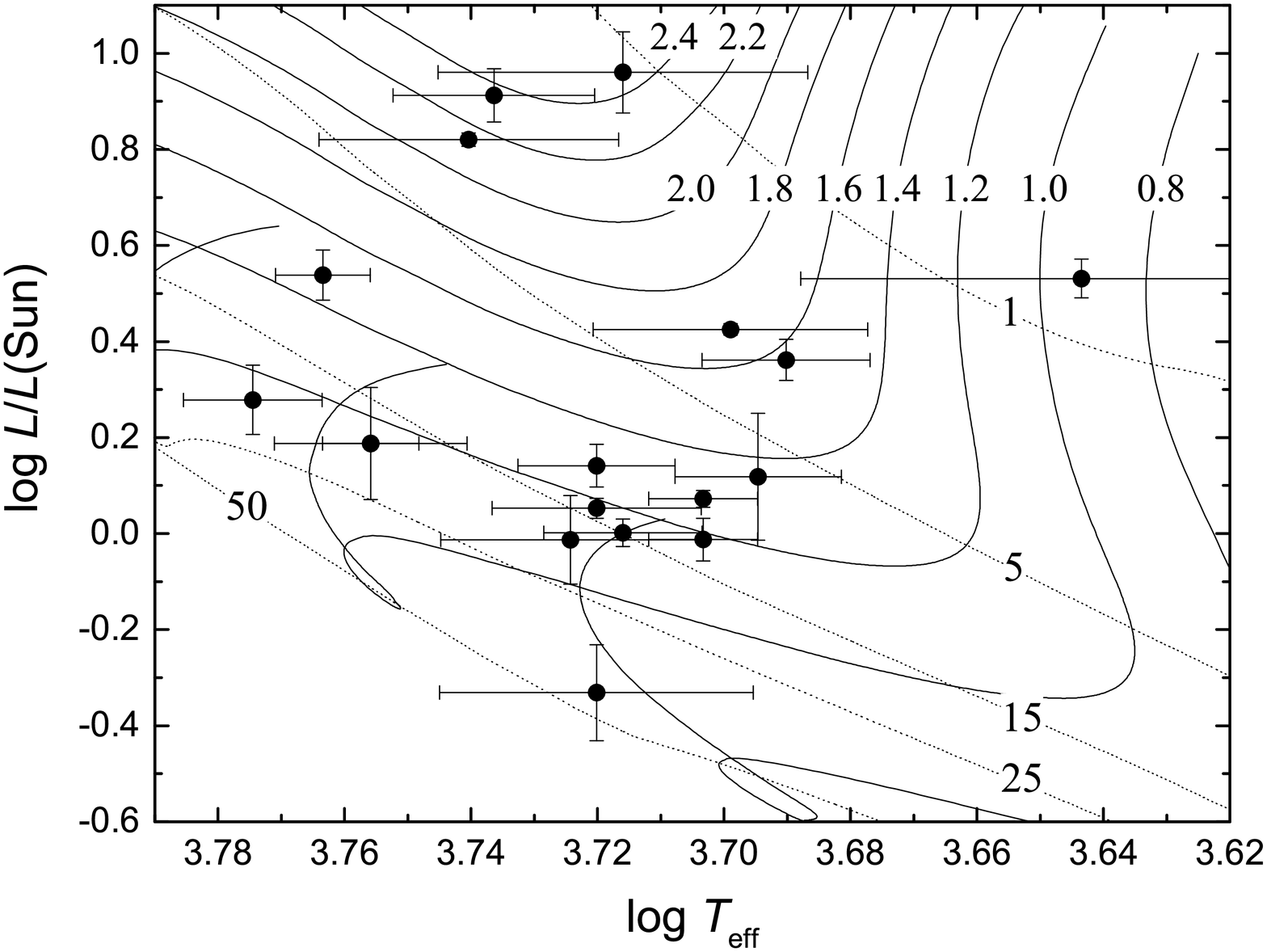} 
  \caption{Location of the target stars in the $\log \Teff$ versus $\log L/\Lsun$\
  diagram together with the PMS evolutionary tracks (solid curves) and isochrones
  (dotted curves) based on the Pisa Stellar Models \citep{tognelli11}.
	The evolutionary tracks range from 2.4 to 0.8\,$\Msun$ and isochrones from 1
  to 50 Myr, as indicated.}
  \label{fig:HRD}
\end{figure*}

Only stars with an available parallax from the \textit{Gaia} DR2 \citep{lindegren18}
or earlier TGAS catalogue \citep{michalik15} 
were included in this analysis. We have used parallax from TGAS only for HD~286178, 
since the value is missing in the newer \textit{Gaia} DR2 catalogue.
However, based on its distance, HD~286178 may not be a member of the Taurus-Auriga SFR.
For HD~283782, the \textit{Gaia} DR2 found two astrometric positions separated by only 1.9~arcsec. 
We have used the parallax of the brighter (in \textit{Gaia} G band) target,
closer to RA$_{2000}$, Dec$_{2000}$ of HD~283782.
No parallax was available for star HD~284135. 

The calculation of luminosity is based on the parallax (or distance), apparent
magnitude, reddening and bolometric correction (BC).
Within the \textit{Gaia} and \textit{Hipparcos} era, we now have quite
accurate parallaxes for stars in the solar vicinity, where also our targets
are located. The estimation of reddening, especially in denser areas such as the
Taurus-Auriga region, still posses several problems and limitations. One way out
of the dilemma could be reddening maps \citep{schlafly14}. However, their
resolutions are not sufficient for our purpose.
We therefore used another approach, by taking available reddening estimates from
the literature \citep{meistas81,chavarria00,grankin13,herczeg14} and the
dereddening method based on the Str{\"o}mgren-Crawford $uvby\beta$ photometric
system \citep{crawford75,schuster89}. The $uvby\beta$ photometric data were
taken from \citet{paunzen15}.
An unweighted mean and its error on $A_{\rm V}$ was calculated (Table \ref{tab:stellar_parameters}).
The values for $A_{\rm V}$ range up to 1.10\,mag with a mean of the error of
0.13\,mag. These values agree with that expected in this region of the sky.
However, the uncertainties are the largest contribution to the overall error in
$\log L/\Lsun$. The BC, especially derived for PMS stars, were taken from \citet{pecaut14}.

The apparent magnitudes are from \citet{grankin13}, except for
HD~30171, HD~283447, HD~283798, and HD~283518 
which are not included in this reference. For these stars, we have used values
from the AAVSO Photometric All-Sky Survey (APASS), after checking the magnitudes
in common with \citet{grankin13}, which yielded an excellent agreement. For
the final error estimation in $\log L/\Lsun$, we applied full propagation of
uncertainties for all the measurements.

A careful assessment of the literature about effective temperature estimation of
T~Tauri stars revealed that there are significant differences in the values for
individual stars. These are caused by the usage of different stellar atmospheres,
spectral resolution, analytic methods and so on. An excellent example of the
challenges to be faced is shown by \citet{lebzelter12} for the case of cool type
Giants. As for reddening, no homogeneous source was found. We therefore
calculated unweighted means of values from the literature
\citep{palla02,ammons06,wright11,grankin13,davies14,mcdonald17} and the 
Str{\"o}mgren-Crawford $uvby\beta$ calibration of \citep{napiwotzki93}.
The latter is based on the $\beta$ index which is an excellent indicator for
the $\log \Teff$. The combination of a narrow and wide filter centred at
H$\beta$ guarantees that any possible emission has no significant effect. The
final uncertainties are between 100 and 450\,K, respectively.
In Table \ref{tab:stellar_parameters} all the derived astrophysical parameters
with their uncertainties are listed.

Finally, we used PMS evolutionary tracks based on the Pisa Stellar Models
\citep{tognelli11} to investigate the evolutionary status of our targets. The
evolutionary tracks with [$X$,$Y$,$Z$] of [0.609,0.2533,0.1377], i.e. solar
metallicity, were used. In Fig. \ref{fig:HRD}, the location of the target stars
in the $\log \Teff$ versus $\log L/\Lsun$\ diagram is shown. The error bars were
used to determine the extent of the possible masses and ages (also in
Table \ref{tab:stellar_parameters}) of candidate stars in comparison to Pisa
Stellar Model isochrones. We have found that the stars in our sample should be
younger than at most 70~Myr. This is close to, or inclusive of, the 10--100~Myr
interval for the post T~Tauri stars (PTTS) defined by \citet{jensen2001}.
The uncertainties, mainly due to effective temperature, allow for the estimating
of masses to within the range $\pm$0.1\,$\Msun$ to $\pm$0.2\,$\Msun$.
To make a comprehensive analysis of the correlation between rotational period
and evolutionary status of our targets, more precise and homogeneous effective
temperatures are required.

\section{Photometric data}
\label{datasets}

For the determination of periods, we needed light curves measured during long
observation campaigns, which are best taken with the same instrument and reduced
with the same pipeline. There are several public data available, notably the
SuperWASP database.
With our access to all of the data presently kept in the SuperWASP archives, we
have based our photometric light-curve investigation on almost 8 years of
observations. The separation into observing seasons is a natural one, since the
target region was best observable from autumn to spring.
To expand our time-domain, we have also added photometry from the NSVS archive.
For ten targets we have found photometry provided by the Kepler \textit{K2} mission.
We break down the individual observing seasons in Table \ref{tab:seasons}.

\subsection{SuperWASP data}
\label{data_swasp}

The WASP instruments have been described by \citet{pollacco06}, and the
reduction techniques discussed by \citet{smalley11} and \citet{holdsworth14}. 
The aperture-extracted photometry from each camera on each night was corrected 
for atmospheric extinction, instrumental colour response, and system zero-point 
relative to a network of local secondary standards. The resulting pseudo-$V$
magnitudes are comparable to {\it Tycho-2\/} \citep{hog00} $V$ magnitudes
\citep{butters10}. In this paper, we have used so far publicly unavailable
SuperWASP data provided by our co-author B.~Smalley.

The SuperWASP data were gathered from July 2004 to January 2012, covering 7
observing seasons in total. The mean season duration is 143 days, however,
season 3 consists of only 12 days. Usually we had five different seasons
available for our targets.
For the star HD~31281 
we found only two seasons of useful data in the archive. The mean cadence of
observations is $\sim$80 seconds.

For all targets, the SuperWASP archive provides 417\,603 points in total and
typically $\sim$4000 points per object in each season.
The SuperWASP pre-clean procedure includes the following steps: 
(i) keeping data with errors of $<$ 0.2 mag, (ii) find the median, and
(iii) keeping data within a brightness interval $\pm$0.2~mag from median.
This procedure reduced our dataset by only 2.4\%.

\subsection{NSVS data}
\label{data_nsvs}

The Northern Sky Variability Survey (NSVS) is a temporal record of the sky over
the optical magnitude range from 8 to 15.5. It was conducted by the
first-generation Robotic Optical Transient Search Experiment (ROTSE-I), using
a robotic system of four co-mounted unfiltered telephoto lenses equipped with
CCD cameras. The survey was conducted from Los Alamos, New Mexico, and primarily
covers the entire northern sky \citep{wozniak04}. The NSVS contains light curves
for approximately 14 million objects with a 1~yr baseline and typically
100--500 measurements per object. The NSVS public data release is available
through a dedicated webpage\footnote{http://skydot.lanl.gov/nsvs/nsvs.php}.

We refer to the NSVS data as ``Season 0'' (in Tables \ref{tab:seasons}, and
\ref{tab:Presult}). We have to note that for HD~31281 
and HD~283572 
no NSVS data are available.
The mean cadence of observations of NSVS data was only $\sim$0.6~d.
Because of few data points ($\sim$105 per star) and the fact that the NSVS data
observed usually only a single point per night, we have opted to not remove any
of them.

To be able to analyse the NSVS data together with the SuperWASP data, we have
made a simple linear transformation in magnitudes. For each target star we have
found the mean magnitude ($m_S$) from the whole pre-cleaned SuperWASP dataset.
Then we have computed the mean magnitude ($m_N$) of the corresponding NSVS
dataset. All the NSVS data were shifted by $m_N-m_S$. We are interested mainly
in temporal position of extremes in the light curves and the possible scaling of
the amplitude of the NSVS light curve is of little concern.

\subsection{Kepler \textit{K2} data}
\label{data_kepler}

The \textit{Kepler Mission} was launched on 2009 March 6. Although it was
designed primarily to detect variable stars and find transiting exoplanets
\citep{Borucki10}, the mission also provided exceptionally high photometric
performance for all stars inside its field of view (FOV). The original FOV was
centred in the constellation Cygnus \citep{Koch10}. 
After pointing and stability problems, the mission was extended to other fields
along the ecliptic and dubbed \textit{K2} \citep{Howell14}. The estimated
photometric precision is down to 400~ppm for stars with $V$\,=\,12\,mag \citep{Howell14}.
In the original proposal, the Taurus-Auriga cluster was planned for the C4
(2015 February-April, centred at R.A.$=03~56~18$, DEC=$+18~39~38$) observational
campaign. After the next extension, another field C13 (2017 March-May, centred
at R.A.$=04~51~11$, DEC=$+20~47~11$) was added\footnote{https://keplerscience.arc.nasa.gov/k2-fields.html}.

Since we were unable to find reliable periods in some observational seasons for
several targets, we wanted to confirm and refine the values by using the
\textit{Kepler} data. With the exception of TTS20, we have only one season of
data per target available. The data were observed with a 30-minute-long exposure
in the Kepler magnitude (white light). Since we were interested in periods longer
than 0.2~day, we have used the provided PDC SAP flux directly corrected for most
systematics by the data conditioning pipeline \citep{stumpe12}.

Because we have used \textit{Kepler} data in a separate analysis, we did not
need to transform the measured fluxes or the Kepler magnitudes.

\begin{table}  
  \begin{center}
  \caption{Observing seasons break-up for the NSVS ($\equiv0$) and
  SuperWASP data. Not all of the target stars were observed in all seasons.
  When using the whole dataset, we referred to the one named $\Sigma$.
  Additional observation by the \textit{K2 Mission} were used as C4 and C13.}
  \label{tab:seasons}
  \begin{tabular}{lccr}
  \hline
  Season   & Start      & End        & Dur. [d] \\
  \hline
  0        & 1999-08-06 & 2000-03-26 & 233      \\
  1        & 2004-07-29 & 2004-09-30 & 63       \\
  2        & 2006-09-17 & 2007-02-27 & 163      \\
  3        & 2008-01-25 & 2008-02-06 & 12       \\
  4        & 2008-10-13 & 2009-03-07 & 145      \\
  5        & 2009-08-12 & 2010-03-26 & 226      \\
  6        & 2010-08-27 & 2011-02-16 & 173      \\
  7        & 2011-09-24 & 2012-01-30 & 128      \\
  $\Sigma$ & 1999-08-06 & 2012-01-30 & 4560     \\
  C4       & 2015-02-08 & 2015-04-20 & 70       \\
  C13      & 2017-03-08 & 2017-05-27 & 80       \\
  \hline 
  \end{tabular}
  \end{center}
\end{table}

\section{Period search}
\label{period}

The majority (70\% of observation nights) of all photometric data was provided
from the SuperWASP archive (seasons 1--7) with mean cadence of 80 seconds.
We have also added ground-based data from the NSVS (season 0) with a mean
cadence over the whole dataset of 0.6~days. However, if there were more than one
data points per observation night the mean cadence was about 45~minutes. Ten out
of the twenty targets were observed in campaign fields of \textit{K2 Mission}
with a 30-minute long cadence. All of the chosen datasets are usable for period
searches in the regime of several days. Although the NSVS dataset has the lowest
cadence, it has the longest duration of a single season (see Table \ref{tab:seasons}).
The previously identified rotational periods were in the range of 0.7 to 3.33~days (Table \ref{tab:basic}).

\begin{figure}
  \centering
  \includegraphics[width=80mm,clip=0]{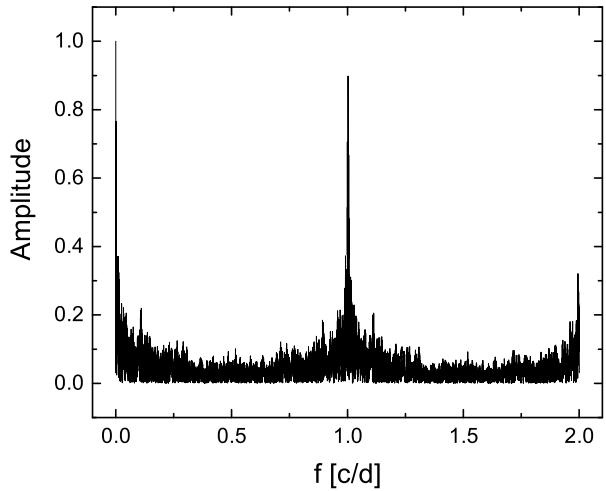}
  \caption{The spectral window of the SuperWASP dataset ($\Sigma$) constructed for HD~285957. 
  This target has the most observation nights available.}
  \label{fig:specwindow}
\end{figure}

We chose to investigate the individual observing seasons separately to account
for possible period changes between them. The problematic season 3 (with only
twelve nights of observation) and season 4 (usually with very few data) were
investigated only when non-trivial Fourier plots were found. Finally, we have
also performed a period search on the combined dataset ($\Sigma$), constituted
from seasons 0--7. The NSVS data were transformed before adding them into the
combined dataset (see section \ref{data_nsvs}). The phased light curves
presented in Figures \ref{fig:phases1} and \ref{fig:phases2} were constructed
using all available ground-based pre-cleaned data points (see section \ref{data_swasp})
and folded on the period corresponding to the most prominent frequency found in
the $\Sigma$ dataset.

\textit{Kepler} datasets (seasons C4 and C13) were investigated separately from
the ground-based data. Except for star HD~285778, only one season of \textit{K2}
data was available per object. In the single special case we have conducted the
period search on both C4 and C13 data separately.

\subsection{Ground-based data}
\label{dcdft}

The period analysis was performed using the Date Compensated Discrete Fourier
Transform algorithm (DC DFT) of \citet{FM81}. The advantage of this method is
that it compensates for gaps within the dataset using weighting, alias
discrimination and harmonic frequency filtering.

The starting frequency was chosen based on the length of the combined dataset
to be $0.00011$~c/d. This corresponds roughly to periods between $0.2$ and
$9000$ days. The step in frequency was set to average $0.0001$~c/d.
Because the NSVS data provided only handful of points per night, we have set
our maximum frequency to 2 cycles per day (c/d) and all Fourier plots
(Figures \ref{fig:periods1} and \ref{fig:periods2}) are calculated up to this
frequency.

We have used two software packages that utilize DC DFT for the search of periods
in our datasets.

The initial analysis was done with the \texttt{VStar} package developed by the
American Association of Variable Star Observers (AAVSO). This tool allowed us to
search for all significant periods (intrinsic or harmonic) and order them with
decreasing power in their Fourier spectrum. It also enabled visual checks to
affirm the periodicity by constructing phase plots for any given frequency (period).

A subsequent analysis was done using the \texttt{Period04} software \citep{lenz04}
with the same frequency range and frequency step for DC DFT.
The code searches only for the most prominent frequency in the Fourier spectrum.
The list of suspected frequencies (periods) from previous investigation using
the \texttt{VStar} package were refined using \texttt{Period04}.
The advantage of \texttt{Period04} was that noise peaks were less likely
to be included in the final list of identified frequencies \citep{breger11}.

To further discern the intrinsic periods from aliases, we have constructed the
spectral window of each dataset (e.g. see Figure \ref{fig:specwindow}).
The most significant period peak was convolved with the spectral window and
subtracted from the previously computed period spectrum (i.e. the corresponding
frequency was pre-whitened for the next analysis).

The result was visually inspected. If the RMS of residuals was too large,
the peak was discarded from the subsequent analysis.
Frequency peaks with a signal-to-noise ratio $>2.0$ were considered for further
analysis. The procedure was repeated until the amplitude of the most significant
peak left was $\le10\%$ of the original most significant period.
Remaining frequencies were included in Table \ref{tab:Presult}.

After periodicity has been identified, we have performed a Marquardt
algorithm-based non-linear least-squares fitting \citep{BR93} simultaneously
with all the identified frequencies, resulting in an adjusted set of parameters
(i.e. frequencies, amplitudes and phases).

The least-squares fit by \texttt{Period04} also provided an error matrix, but
the derived uncertainties only estimate the consistency of fit.
The \texttt{Period04} package has a tool to compute more realistic uncertainties
using Monte-Carlo simulations, which we have used with the default values.

The detected periodic signals were subtracted out and the remaining data points
were randomly rearranged with the original timestamps preserved.
A pseudo-random (based on the on-board computer time) Gaussian noise with
a standard deviation of 0.1 was added to the magnitudes predicted by the last fit.
For each optimization a set of total 100 test time strings was generated.
The identification of peaks and least-square fit to the light curve was carried
out for each of the simulated time strings. The uncertainties were calculated on
the distribution of fit parameters \citep{lenz05}.

\subsection{\textit{Kepler} data}
\label{K2period}

The period analysis was performed separately from the ground-based data.
However, the same frequency domain (0.00011--2~c/d) was searched with a step of
$\Delta f=0.0001$~c/d. The mean data error was only 26~ppm. Combined
with the 30-minute cadence and almost uninterrupted 70--80 days of observation,
the resulting Fourier plots (Figure \ref{fig:periodsK2}) are without much noise
and period peaks are easily distinguishable. The peaks of most the significant
periods in the regime of several days are corresponding to intrinsic changes of
the brightness and no spectral window aliases were expected above the frequency
of $\sim1/80=0.0125$~c/d.

The resulting most significant periods are presented in Table~\ref{tab:K2periods}.
We have compared the periods found from the \textit{Kepler} data with the mean
periods found in our previous analysis of SuperWASP and NSVS data, as well as
with the values found in the literature. 

The period analysis of \textit{Kepler} data yielded similar periods to those
obtained from the ground-based data only. The mean difference was of the order
of $\sim0.1$~days. In comparison to Table~\ref{tab:Presult}, the mean
uncertainties in periods were smaller because of clear and prominent peaks in
Fourier plots. Several dubious periods (e.g. for BD+19~656) 
found in some observing seasons in SuperWASP data were also discovered with
smaller amplitudes, and corresponded to a near 1:2 period ratio with the most
prominent period. This indicates that spots areas are
distributed along the length of the equator. For stars HD~283782 
and HD~31281 
we were able to find first reliable periods only from the \textit{Kepler} data.
Objects with different period estimates are discussed further in
Section~\ref{discus} below.

In addition to period determination, we have looked at the possible evolution of
the period during the $\sim$80-day long \textit{Kepler} observation. We have
used the Wavelet Z-transform (WWZ) algorithm by \citet{Fos96}, provided within
the \texttt{VStar} package. The algorithm can achieve resolution in both
frequency and time for unevenly spaced data. The wavelet was a sinusoidal wave
shifted with a constant term. A sliding window of a constant width (in time)
moved across the data.

The highest weight was applied to the data points closest to the centre of the
window. The output was a relation between the signal power (Z value) as function
of time and frequency. We were interested in tracking the maximum Z value for
any given time. The frequency corresponding to the maximum Z value was
interpreted as 1/$P$ (see e.g. Figure~\ref{fig:TTS15WWZ}).

To cover the period range from all investigated targets, we have run the
wavelet analysis for periods from $0.5$ to $4.0$ days with the light curve
separated into 50 time bins. The step in period was set to $\Delta P=0.01$~days
(half of the cadence interval) and the so-called decay parameter was fixed
at the default value 0.001. The maximum Z-values for individual time bins were
plotted along with the corresponding semi-amplitudes of the light curve
(see Figure~\ref{fig:K2wwz}). Since we had only the total flux in the \textit{Kepler}
passband, we have expressed the amplitude of the light curve only as a ratio
and not in magnitudes.

We have also constructed a simple spot model to compute artificial light curves
with emerging and disappearing spots. The spherical stellar surface was covered
with 360$\times$181 mesh in longitude and latitude. Flux was calculated for each
grid point taking into account the size of the local grid, the presence of spots
and visibility for the observer. A linear limb-darkening law was used.
Spots were placed at arbitrary longitudes (on the equator) and revolved with the
rotation period. The inclination of stellar rotation axis was zero and a rigid
rotation was assumed.
We were able to set the time of appearance and disappearance of individual spots.

\begin{figure*}
  \centering
  \includegraphics[width=161mm,clip=0]{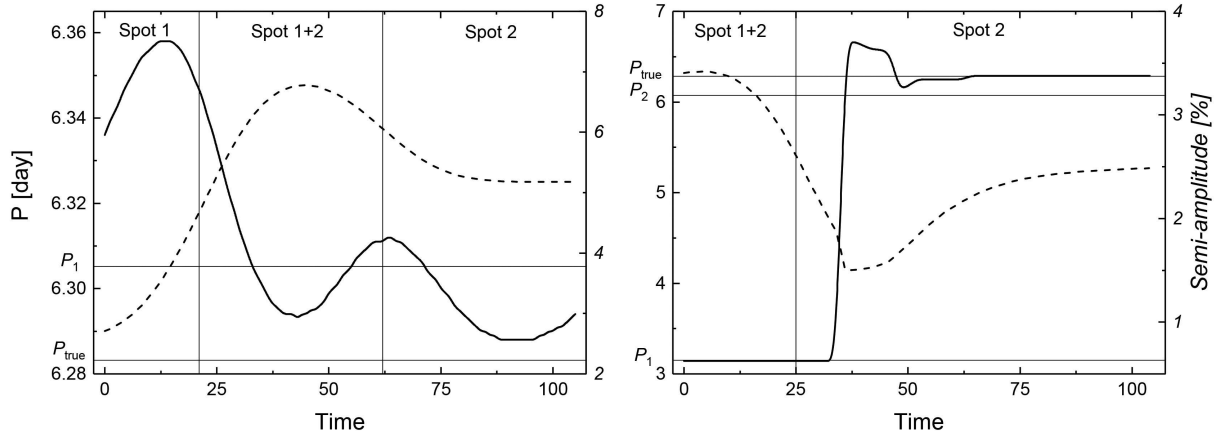}
  \caption{The evolution of period (solid line) and amplitude (dashed line) of
  the simple spot model light curve.   The left panel shows the effect of a new
  randomly-placed spot and subsequent disappearance of the original spot.
  The right panel shows the effect of two spots placed on the opposite sides of
  the stellar surface with later disappearance of the original spot.}
  \label{fig:modelWWZ}
\end{figure*}

We have generated long (duration $>10 P_{\rm rot}$) light curves for several
scenarios and applied both period and wavelet analyses using
the \texttt{VStar} package. Results of two different setups are presented in
Figure~\ref{fig:modelWWZ}.

\begingroup
\setlength{\tabcolsep}{2pt}
\begin{landscape}
\begin{table}  
  \begin{center}
  \caption{Results of period analysis for all target objects.
  The prominent periods for the same object listed by decreasing power.
  Seasons without any usable data are marked as ``--''. The last 
  column lists periods obtained from the literature (see Table~\ref{tab:basic})
  ``?'' denotes uncertain values. Values in brackets ``$\langle$'',  ``$\rangle$''
  are not considered as the intrinsic period.}
  \label{tab:Presult}
  \small
  \begin{tabular}{lccccccccccc}
  \hline
  Object            & \multicolumn{8}{c}{Most prominent period [d] in seasons}                                                                                                                                                              & $P_{\rm mean}$ [d]                       & $A_{\rm mean}$ [mag] & $P_{\rm lit}$ [d]      \\
                    & 0                        & 1                        & 2                        & 3                        & 4                        & 5                        & 6                        & 7                        & $\Sigma$                                 &                      &                        \\ 
  \hline 
  HD~285281         & 1.1697(37)               & 1.1689(29)               & 1.1705(39)               & --                       & --                       & 1.1706(44)               & 1.1722(36)               & 1.1716(12)               & 1.1711(37)                               & 0.0413(51)           & 1.1683                 \\
  \tvspc 
  BD+19~656         &$\langle$1.0041(285)$\rangle$&$\langle$1.4421(207)$\rangle$& 2.9104(353)        & --                       & --                       & 2.8885(59)               & 2.8869(289)              &$\langle$1.4611(188)$\rangle$& 2.8849(51)                            & 0.0080(3)            & 2.8600                 \\
                    &                          &                          &                          &                          &                          &                          &                          &                          &                                          &                      &$\langle$0.7410$\rangle$\\
  \tvspc 
  HD~284135         &$\langle$0.9986(204)$\rangle$& 0.8181(144)           & 0.8183(69)               & --                       & --                       & 0.8175(325)              & 0.9175(121)              & 0.8245(67)               & 0.8179(58)                               & 0.0106(20)           & 0.8160                 \\
  \tvspc 
  HD~284149         & 1.0309(316)              & 1.0353(192)              & 1.0534(224)              & --                       & --                       & 1.0473(268)              & 1.0419(362)              & 1.0684(84)               & 1.0712(7)                                & 0.0084(3)            & 1.0790                 \\
  \tvspc 
  HD~281691         & 2.6596(171)              & 2.6511(77)               & 2.6610(294)              & --                       & --                       & 2.6274(339)              & 2.6589(328)              & 2.6226(157)              & 2.6267(237)                              & 0.0177(17)           & 2.74?                  \\
  \tvspc 
  HD~284266         & 1.8086(185)              & 1.8218(237)              & 1.8087(183)              & --                       & --                       &$\langle$0.9016(133)$\rangle$& 1.8396(24)            & 1.8403(88)               & 1.8433(10)                               & 0.0315(4)            & 1.83                   \\
  \tvspc 
  HD~284503         & 0.5166(183)              & 0.7396(148)              & 0.7369(227)              & --                       & --                       & 0.7308(78)               & 0.7363(64)               & 0.7369(58)               & 0.7370(3)                                & 0.0267(25)           & 0.736                  \\
  \tvspc 
  HD~284496         & 2.7248(211)              & 2.7056(205)              & 2.7390(336)              & --                       & --                       & 2.6617(266)              & 2.7556(882)              & 2.6846(763)              & 2.6880(195)                              & 0.0486(23)           & 2.71                   \\
  \tvspc 
  HD~285840         & 1.5562(177)              & --                       &$\langle$1.2241(3247)$\rangle$& 1.5323(1134)         & --                       & 1.5470(395)              & 1.5751(306)              & 1.5538(224)              & 1.5476(67)                               & 0.0315(26)           & 1.55                   \\
  \tvspc 
  HD~285957         & 3.0874(294)              & --                       & 3.0497(330)              &$\langle$4.1736(5446)$\rangle$& 3.0798(236)          & 3.0950(534)              & 3.0619(383)              & 3.0931(336)              & 3.0546(255)                              & 0.0251(15)           & 3.07                   \\
  \tvspc 
  HD~283798         & 0.9868(107)              &$\langle$0.8906(1934)$\rangle$& 0.9861(92)           & --                       & --                       &$\langle$0.9957(340)$\rangle$& 0.9890(154)           & 0.9860(93)               & 0.9872(33)                               & 0.0159(5)            & 0.6?                   \\
  \tvspc 
  HD~283782         &$\langle$1.0016(91)$\rangle$&$\langle$1.0413(319)$\rangle$&$\langle$0.8706(1873)$\rangle$& --              & --                       &$\langle$1.0589(872)$\rangle$&$\langle$0.8716(1333)$\rangle$&$\langle$0.8684(1139)$\rangle$&$\langle$0.8704(1106)$\rangle$& 0.0081(26)            & ?                      \\
  \tvspc 
  HD~30171          &$\langle$0.9956(2272)$\rangle$& --                   & 1.1053(873)              & 1.1013(1057)             & 1.1106(141)              & 1.1055(210)              & 1.1101(141)              & 1.1062(164)              & 1.1058(33)                               & 0.0272(13)           & 1.104                  \\
                    &                          &                          &                          &                          &$\langle$0.5254(32)$\rangle$&                        &                          &                          &                                          &                      &                        \\
  \tvspc 
  HD~31281          & --                       & --                       &$\langle$0.7920(1100)$\rangle$&$\langle$0.7960(1181)$\rangle$& --               & --                       & --                       & --                       &$\langle$0.7913(15)$\rangle$              & 0.0098(12)           & ?                      \\
  \tvspc 
  HD~286179         & 3.2573(627)              & --                       & 3.1486(1914)             &$\langle$0.9272(621)$\rangle$& 3.1260(401)           & 3.1476(501)              & 3.3201(199)              & --                       & 3.1397(221)                              & 0.0294(13)           & 3.33                   \\
  \tvspc 
  HD~286178         &$\langle$0.7490(15)$\rangle$& --                     &$\langle$1.6958(281)$\rangle$&$\langle$0.7759(1023)$\rangle$& 2.3680(399)       &$\langle$1.6981(329)$\rangle$&$\langle$1.6990(387)$\rangle$& --                 &$\langle$1.7001(81)$\rangle$              & 0.0231(37)           & 2.39                   \\
                    & 2.0442(125)              &                          & 2.4242(551)              & 2.0198(210)              &$\langle$1.7013(220)$\rangle$& 2.4155(642)           & 2.4155(735)              &                          & 2.4125(164)                              & 0.0227(34)           &                        \\
  \tvspc 
  HD~283447         & 3.0883(239)              & 3.0741(1071)             & 3.0912(268)              & --                       & --                       & 3.0628(141)              & --                       & 3.0826(374)              & 3.0836(210)                              & 0.0695(24)           & 51                     \\
  \tvspc 
  HD~283572         & --                       & 1.5392(175)              & 1.5475(227)              & --                       & --                       &$\langle$1.4077(829)$\rangle$& 1.5823(544)           & 1.5470(208)              & 1.5462(38)                               & 0.0386(22)           & 1.529                  \\
  \tvspc 
  HD~285778         &$\langle$0.9951(333)$\rangle$& --                    &$\langle$1.2602(194)$\rangle$&$\langle$0.9801(879)$\rangle$& 2.7293(1542)       & 2.7412(541)              & 2.7360(1030)             & 2.7308(759)              & 2.7361(204)                              & 0.0132(32)           & 2.734                  \\
  \tvspc 
  HD~283518         & 1.8723(26)               & 1.8716(63)               & 1.8713(645)              & --                       & --                       & 1.8734(641)              &$\langle$0.9361(118)$\rangle$& 1.8702(377)           & 1.8706(14)                               & 0.0491(18)           & 1.87                   \\
                    &$\langle$0.9362(27)$\rangle$&$\langle$0.9358(35)$\rangle$&$\langle$0.9356(151)$\rangle$&                   &                          &$\langle$0.9367(239)$\rangle$&                       &$\langle$0.9351(92)$\rangle$&                                        &                      &                        \\
  \hline
  \multicolumn{12}{l}{Note: Uncertain periods are further discussed in section \ref{discus}.}\\
  \end{tabular}
  \end{center}
\end{table}
\end{landscape}
\endgroup

One typical situation similar to our Sun is when a new spot emerges on the
stellar limb while the original spot is still present. After several days, the
older spot declines and disappears while the
new one is still present (see the left panel in Figure~\ref{fig:modelWWZ}).
The Fourier analysis will find only a ``median'' period $P_1$ (of periods in
the individual spot evolution time step).

\begingroup
\setlength{\tabcolsep}{5pt}
\begin{table}
  \begin{center}
  \caption{Results of period analysis for objects with available \textit{Kepler}
  data. Periods, $P_{\rm Kepler}$, are ordered by decreasing power of the
  corresponding peak. Values in brackets ``$\langle$'', ``$\rangle$'' are not
  considered as the intrinsic period. Comparison with previously determined
  periods ($P_{\rm lit}$) and periods based on SuperWASP data ($P_{\rm mean}$)
  are also shown.}
  \label{tab:K2periods}
  \begin{tabular}{lcccc}
  \hline
  Object    & $P_{\rm lit}$ [d] & $P_{\rm mean}$ [d]             & $P_{\rm Kepler}$ [d]             & Field \\
  \hline
  BD+19~656 & 2.86              & 2.8849(51)                     & 2.8489(8)                        & C4    \\ 
            &                   &                                & $\langle$1.4467(4)$\rangle^*$    &       \\ 
  \tvspc
  HD~284496 & 2.71              & 2.6880(195)                    & 2.7738(8)                        & C13   \\ 
            &                   &                                & $\langle$2.6525(6)$\rangle$      &       \\ 
  \tvspc
  HD~285840 & 1.55              & 1.5476(67)                     & 1.5463(2)                        & C13   \\ 
            &                   &                                & $\langle$0.7735(1)$\rangle^*$    &       \\ 
  \tvspc
  HD~285957 & 3.07              & 3.0546(255)                    & 3.0863(10)                       & C13   \\ 
            &                   &                                & $\langle$1.5311(2)$\rangle^*$    &       \\ 
  \tvspc
  HD~283798 & 0.6?              & 0.9872(33)                     & 0.9831(2)                        & C13   \\ 
            &                   &                                & $\langle$0.9658(2)$\rangle$      &       \\ 
  \tvspc
  HD~283782 & ?                 & $\langle$0.8704(1106)$\rangle$ & 2.0181(4)                        & C13   \\ 
  \tvspc
  HD~31281  & ?                 & $\langle$0.7913(15)$\rangle$   & 0.6771(1)                        & C13   \\ 
            &                   &                                & $\langle$0.7999(1)$\rangle$      &       \\ 
  \tvspc
  HD~286179 & 3.33              & 3.1397(221)                    & 3.1249(20)                       & C13   \\ 
            &                   &                                & $\langle$1.5647(5)$\rangle^*$    &       \\ 
  \tvspc
  HD~286178 & 2.39              & 1.7001(81)                     & 1.7027(6)                        & C13   \\ 
            &                   &                                & $\langle$2.3562(11)$\rangle^*$   &       \\ 
            &                   &                                & $\langle$1.1813(3)$\rangle$!     &       \\ 
  \tvspc
  HD~285778 & 2.734             & 2.7361(204)                    & 2.8554(17)                       & C4    \\ 
            &                   &                                & $\langle$1.3717(4)$\rangle^*$    &       \\ 
  \tvspc
            &                   &                                & 2.7510(15)                       & C13   \\ 
  \hline
  \multicolumn{5}{l}{Note: Periods marked with ``$^*$'' are close to 1:2 ratio with}\\
  \multicolumn{5}{l}{the more prominent period. See the discussion in text for}\\
	\multicolumn{5}{l}{details on special cases marked with ``!''.}\\
  \end{tabular}
  \end{center}
\end{table}
\endgroup

If two spotted regions were placed almost on the opposite sides of the stellar
surface a sudden change of period is encountered (see the right panel in
Figure~\ref{fig:modelWWZ}). Fourier analysis finds two most significant periods
($P_1$, $P_2$). The intrinsic rotation period was closer to the longer period
found by Fourier analysis. This means that we have to be cautious about
interpreting most significant periods of \textit{Kepler} targets. The evolution
of rotation period and its amplitude during $\sim$80 days for all target stars
is shown in Figure~\ref{fig:K2wwz}.

\subsection{Discussion on individual targets} 
\label{discus}

The adopted periods we discuss here stem from the period analysis on the whole
ground-based dataset ($\Sigma$). We include a note if we have found different
non-harmonic periods in a singular observing seasons (see Table \ref{tab:Presult}).
Wherever possible, we have preferred the period found in the \textit{Kepler}
dataset (Table \ref{tab:K2periods}). Fourier plots and wavelet analysis of
period evolution can be found in the appendix (Figures~\ref{fig:periodsK2}
and \ref{fig:K2wwz}).

All periods found for the objects in our sample are interpreted as rotation
periods due to magnetic cool spots on the surface of the star. There are other
possible sources of periodic signals in T~Tauri stars, e.g. hot spots in the
circumstellar disk, and/or obscuration with material at the inner disk edge.
However, studies using the IRAS \citep{strom89} and Spitzer \citep{padgett06}
telescopes found that such disks are rare among WTTS.
\citet{hartmann05} found that in the Taurus-Auriga region, the CTTS
are clearly separated form WTTS based on the infrared colour excess.
So-far there have been no disks detected around stars in our sample.

For star \textbf{BD+19~656} 
we have found two periods in our ground-based
data analysed season-wise. However, the main period was found when analysing the
whole data set and also the \textit{K2} data to be $\sim2.85$ days, which
corresponds to the value previously reported in literature \citep[see e.g.][]{grankin13}.
Based on the modelled period evolution, the intrinsic rotation period may be shorter.
We have visually examined the \textit{Kepler} light curve phased with both
prominent test periods separately (see Figure~\ref{fig:K2phased}) and decided to
adopt the period $P=2.8489$ days.

Periods of \textbf{HD~284496} 
in different observational seasons ranged
from $\sim2.66$ to $\sim2.76$ days. These two limits were resolved as separate
close peaks by the period analysis of \textit{Kepler} data (see Figure~\ref{fig:periodsK2}).
The most prominent period $P=2.7738(8)$ days was found to be the central value
of period changes during the C13 season, investigated by the wavelet analysis.
Based on the modelled period evolution, the intrinsic rotation period may be shorter.

For stars \textbf{HD~285840} 
and \textbf{HD~285957} 
the period analysis of the ground-based data and \textit{Kepler} data provided
the same results. The outlier periods found in a single season do not have any
physical meaning.
Also, the wavelet analysis showed no significant changes ($\sim0.02$\,days)
over the range of 80\,days. The second most significant period found in the \textit{K2}
data is in 1:2 ratio to the mean period. This could indicate that some of the
larger spotted regions are located on the opposite sides of the stellar surface.
However, the modelled period evolution showed no sudden change in period. This
could be explained by e.g. the both spotted regions are present on the surface
and small changes of period can be attributed to differential rotation of spots
changing their latitude. However, by folding the \textit{Kepler} light curve
with longer and shorter test periods for both stars (see Figure~\ref{fig:K2phased}),
we have adopted the periods $P=1.5463$~days and $P=3.0863$~days for HD~285840
and HD~285957, respectively.

The star \textbf{HD~283798} 
was found to have a period of $P\sim0.987$ days. However, the corresponding
frequency ($f=1.013$\,c/d) was distinguished from the peak
at $f=1.0$ (see Figure~\ref{fig:periods2}) and with higher amplitude. This is an
unfavourable case for ground-base observations and could be the main reason why
the star had no published period in the literature. However, in the \textit{K2}
dataset we have found clear peaks at $P=0.9831(2)$ and $P=0.9658(2)$\,days.

We have found a previously unpublished period $P=0.87(11)$~days for \textbf{HD~283782}. 
However, in several observing seasons we have detected periods close to one day.
The Fourier plot (see Figure~\ref{fig:periods2}) was noisy with the mean SNR of 1.7.
Because of large uncertainties and spurious periods, we have performed a
separate period analysis on the \textit{Kepler} data.
We have found a different prominent peak at $P=2.0181(4)$ days. The amplitude of
the \textit{Kepler} light curve is only about $0.25\%$. Other period peaks were
present only in the red-noise region of the Fourier plot (see Figure~\ref{fig:periodsK2}).
We have checked the data distribution in SuperWASP seasons 2, 6, and 7
(i.e. season which yielded the period $\sim0.87$~day). The mean data cadence
was $\sim72$ seconds ($\sim0.0008$ day).
We have investigated possible harmonic periods from the data sampling in these
seasons. The length of uninterrupted observation of the star per night was found
to be close to a normal distribution with the centre at $\sim4.6$~hours ($0.192$~day).
The corresponding frequency is above our threshold 4.8~c/d. A new period search
extended up to 20~c/d was performed on the whole ground-base dataset.
Interestingly, a new most prominent period of 0.317~days (3.1545~c/d) was found.
The amplitude of frequency corresponding to period 0.192~day was less than $10\%$
of the new most prominent period. We have not been able to link the periods
0.192~day, 0.317~day, or their harmonic multiples to any other period peak found
in the Fourier plots for seasons 2, 6, and 7.
However, the same period analysis on the \textit{K2} data arrived to the most
prominent period $P=2.0181(4)$ days. Based on the modelled period evolution, the
intrinsic rotation period may be shorter. We have adopted the \textit{Kepler}
period and consider all estimates from the ground-based data (in Table \ref{tab:Presult})
to suffer from sampling effects.
We have found first reliable period for \textbf{HD~283782} 
using the \textit{Kepler} dataset. However, the evolution of this period
computed by WWZ shows a decline from mean by $\sim$0.2~days accompanied by
a small drop in amplitude towards the end of continuous observation.
A longer dataset would be needed to investigate further. Also, the mean period
2.02~days also provides problems for ground-based monitoring.

\begin{figure}
  \centering
  \includegraphics[width=80mm,clip=0]{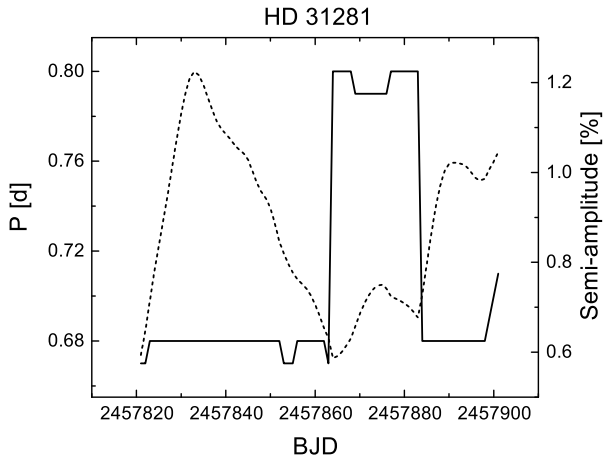}
  \caption{The evolution of period (solid line) and amplitude (dashed line)
  of the light curve of HD~31281 
  during the C4 campaign of \textit{K2 Mission}. See text for details.}
  \label{fig:TTS15WWZ}
\end{figure}

Analysing the dataset for \textbf{HD~31281} 
is not straightforward because it was observed only in seasons 2 and 3.
The mean cadence was $\sim$300~seconds spread on only 74 days of observations.
This resulted into a high amplitude of the diurnal frequency ($f=1$) and also the
red-noise contribution was substantial. The intrinsic frequency was found only
as the third most significant one (see peaks marked with circles in Figure~\ref{fig:periods2}).
Two of the most significant periods were
pre-whitened leaving us with $P\sim0.79$~day from the analysis of all data.
The found amplitude was also small $\sim0.01$~mag, which is probably the reason
no period was published so far for this object. Fortunately, the \textit{K2} data
provided a mean period $P=0.6771(1)$~day. The WWZ analysis showed two distinct periods
present in the course of 80 days. A higher period ($\sim0.8$~day) was present for
about 1/4 of the time interval. This corresponds to season 3 of SuperWASP data.
The sudden change from 0.68~day to 0.79~day (lasting for about 20~days) can be
interpreted with the emergence of another cool spotted region at the surface of
the star together with the disappearance of a previous spotted region on the
eastern limb of the star. If the change would be close to 50\% we could argue
that the spotted regions were almost at opposing sides. This was further
accompanied by the change of amplitude of the light curve
(see Figure~\ref{fig:TTS15WWZ}). During the period-change event the amplitude
curve has local minima, e.g. the percentage of surface covered with cool spots
is higher at these times.
We adopted the lower period as the intrinsic one. Based on the modelled
period evolution, the intrinsic rotation period may be shorter.

The analysis of ground-based observation by SuperWASP and NSVS and \textit{Kepler}
data yielded the same result for star \textbf{HD~286179}, 
namely $P\sim3.1397(21)$~days and
$P\sim3.1249(20)$~days, respectively. This value is smaller than previously reported
by \citet{bouvier97} (3.33 days). A harmonic 1:2 period peak was also found in
the C13 dataset. We have folded the \textit{Kepler} light curve using both test
periods (Figure~\ref{fig:K2phased}). After visually inspecting the resulting
phase curve, we have adopted the period $P=3.1249$ days. The origin of period
$P=0.9272(621)$~days in Season 3 of SuperWASP is likely the result of a smaller
number of data in comparison to other seasons.

\begin{figure}
  \centering
  \includegraphics[width=80mm,clip=0]{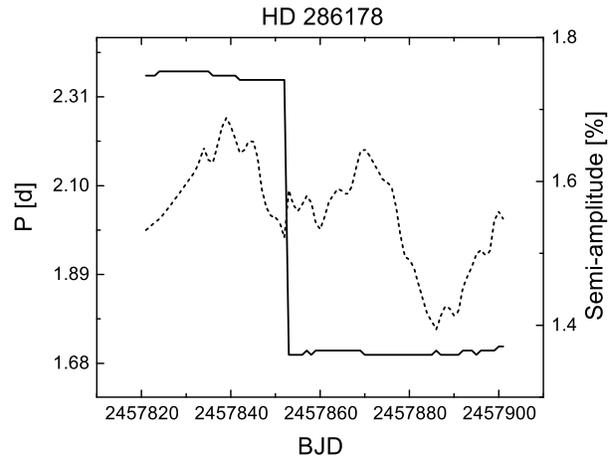}
  \caption{The evolution of period (solid line) and amplitude (dashed line) of
  the light curve of HD~286178 
  during the C13 campaign of \textit{K2 Mission}. See text for details.}
  \label{fig:TTS17WWZ}
\end{figure}

When analysing data of \textbf{HD~286178} 
we found the most prominent period $P\sim1.7$~days.
This is the 1:2 harmonic of $P=2.39$~days reported by \citet{bouvier97}.
However, in our season 4 (2008--2009), we have found the most prominent
period to be $\sim2.37$~days. Data of \citet{bouvier97} were obtained during
years 1994--1996. The change of the observed period can be explained, e.g. by
emerging a previously unobserved area of photospheric spots.
This would result in the shorter photometric period of $\sim1.7$~days.
We have repeated the period analysis with all the ground-based data binned to a
single point per night spanning 12.5~years. The subsequent period analysis showed
only one peak at $P\sim2.4$~days. We are thus inclined to consider the rotation
period to be close to this value. The \textit{Kepler} dataset revealed prominent
peaks at $P=1.7027(6)$, $2.3562(11)$, and $1.1813(3)$~days (ordered with
decreasing power). Also the wavelet analysis showed a sudden change of period
from $\sim2.36$ to $\sim1.70$~days (see Figure~\ref{fig:TTS17WWZ}).
Based on the distance of 74.3~pc of HD~286178 (from Table~\ref{tab:basic}) and
stellar properties of $R=1.04$~$\Rsun$, $L=0.71$~$\Lsun$, and a reliable
estimate of $A_{\rm V}$ we can write
\[R\sin i = (P_{\rm rot} v\sin i)/2\pi.\]
If we use the value of $v\sin i=42$~km.s$^{-1}$ by \citet{wichmann00}, we obtain
the values for $\sin i$. However, only for the period $P=1.1813(3)$~days is
$\sin i\leq1$ which disallows the other periods $P=1.7027(6)$~days
and $2.3562(11)$~days from further analysis.

\textbf{HD~283447} 
had a previously reported period $P\sim51$~days \citep{welty95}.
The star was described as an SB2 system with K-type components. From the photometric
point of view we have not found any signal corresponding to this period, i.e.
the measured variations in flux are due to the changes on the visible photosphere.
We have found a period $P=3.0836(210)$~days with an amplitude $\sim2\times$ the
diurnal period. 

\textbf{HD~285778} 
was found to have a mean period of $\sim2.7361(204)$~days
from the ground-based data, which confirms the previously determined period by
\citet{grankin13}. This object was also the only one from our sample to have
two \textit{Kepler} datasets available. In the C4 season an additional period
was found. We have folded the data using both test periods (the result is shown
in Figure~\ref{fig:K2phased}). The descending slope between phases 0.2--0.8
folded with the period $P=2.8554$~days is closer to the descending slope of the
original light curve. The dubious period from the NSVS data $0.9951(333)$~days
is most probably a window alias. The determination of a $0.9801(879)$~day period
from season 3 was plagued by the low number of data points. In season 2, the
period found was close to 1:2 and is caused by spots distributed by $\sim180$
degrees on the surface of the star.

\begin{figure}
  \centering
  \includegraphics[width=80mm,clip=0]{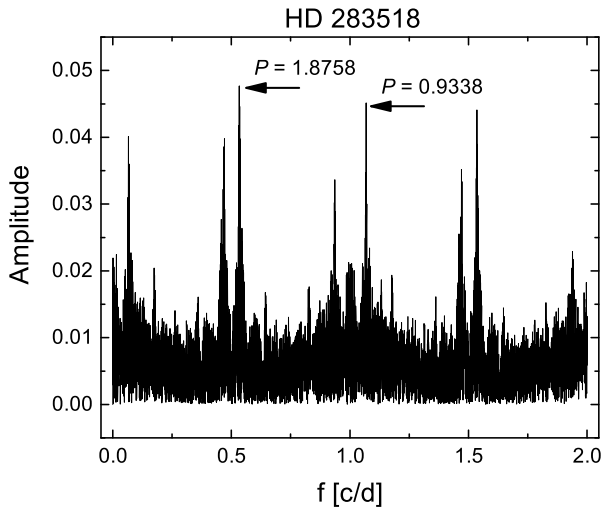}
  \caption{Fourier plot of the whole dataset ($\Sigma$) for HD~283518. 
  Most prominent periods are marked. Other peaks are aliases from the spectral window.}
  \label{fig:TTS21Period}
\end{figure}

\begin{figure}
  \centering
  \includegraphics[width=80mm,clip=0]{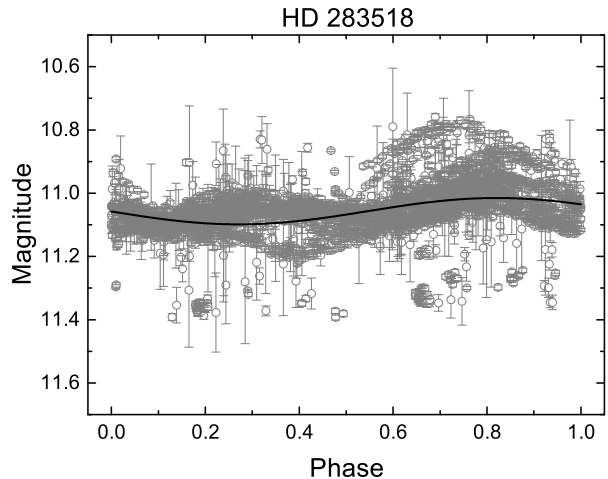}
  \caption{Light curve of all available photometry from SuperWASP and NSVS data
  for the HD~283518 
  folded with the mean period $P=1.8706$ days.} 
  \label{fig:TTS21PhaseLC}
\end{figure}

\begin{figure*}
  \centering
  \includegraphics[width=159mm,clip=0]{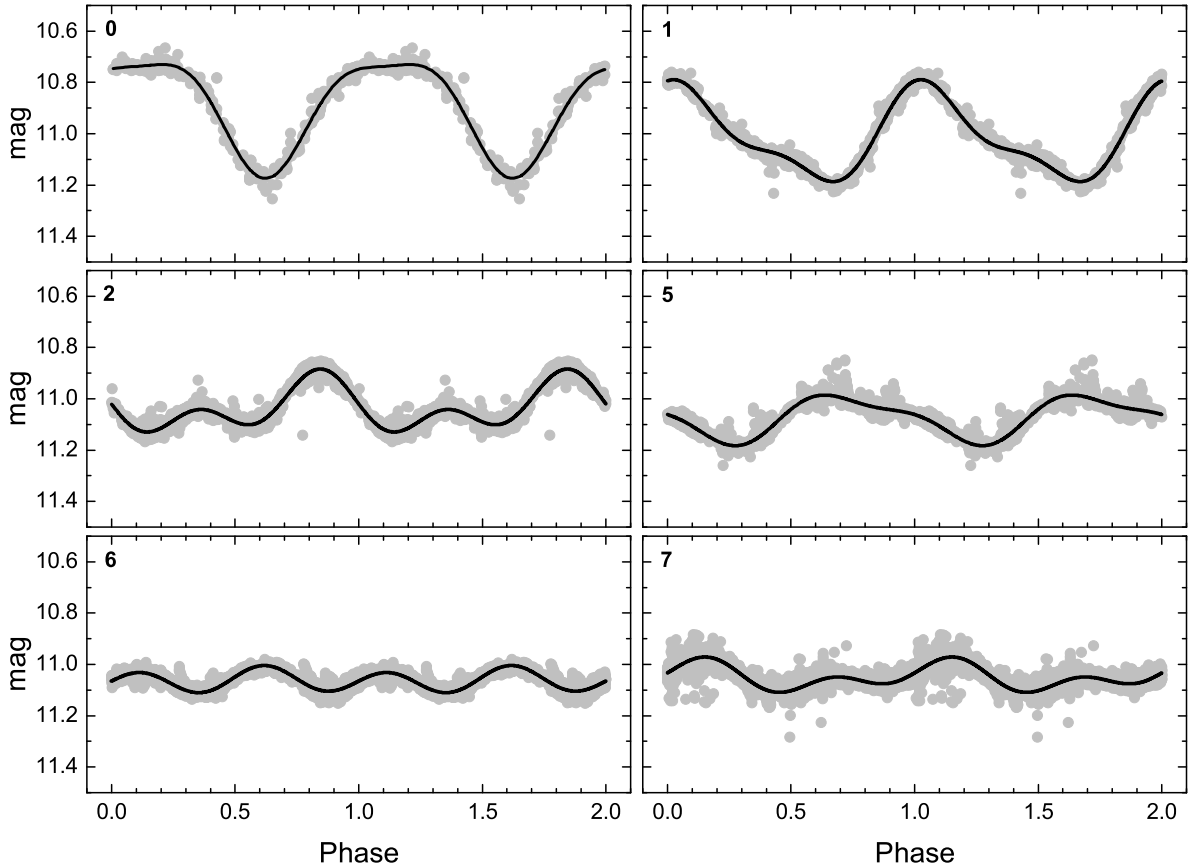}
  \caption{Evolution of the phased light curve of star HD~283518 
  in different observational seasons (panel number). The most dominant frequency
  in each season (see Table~\ref{tab:Presult}) was used to construct the phased
  light curve. The solid line represents a fit with the addition of the first
  harmonic frequency.}
  \label{fig:fits21}
\end{figure*}

The most intriguing object is \textbf{HD~283518} (V410~Tau). 
This is a well-investigated star \cite[see e.g.][]{stelzer03,gran09}.
The phase light curve of the whole dataset ($\Sigma$)
shows a clear frequency wave (see Figure~\ref{fig:TTS21PhaseLC}) when folded
with the most prominent period. The period analysis based on the whole dataset
yielded two distinct peaks (Figure~\ref{fig:TTS21Period}) at
$P=1.8758$ days, and $P=0.9338$ days (ordered by their respective amplitude).
We have to apply the DC DFT analysis for separate seasons
(0--2 and 5--7, see Table~\ref{tab:seasons} for dates of observations).
Except for season 6, we have found the mean
period $\sim1.87$~day. To properly fit the light curve, we had to add the next
harmonic frequency corresponding roughly to $P_2\sim0.936$~days. Phased light
curves with a fit are presented in Figure~\ref{fig:fits21}. The added harmonic
period can be interpreted as due to the presence of at least another spot
located almost opposite the first spot area. Detailed spot configuration would
need to be carried out by a Zeeman-Doppler imaging and surface inversion technique.

\section{Conclusions}
\label{final}

We have performed a thorough period analysis of twenty selected weak-line
T~Tauri stars based on the data available from the NSVS and the SuperWASP archive.
For two objects (HD~283782 and HD~31281) 
we made the first estimate of the photometric period.
Also we have found a more reliable period estimate for HD~283798. 
Period analysis based on all available data lead to the confirmation and update
of periods previously published in literature in 15 cases (up to 10\% difference).

The main advantage of these period estimates is that we have analysed a very
long time interval (up to $\sim$12.5~years). Fortunately, our targets with
dubious periods were also observed by the \textit{K2 Mission}, which allowed us
to distinguish the mean photometric periods from their harmonic components.
We also have performed wavelet analysis for our stars observed by \textit{Kepler}
and searched for period evolution. The sudden change of period in several
targets can be explained by the emergence of another photospheric spot which
remained visible for several days.

We have constructed a simple spot model with the possibility of adding and
removing spots to compute long artificial light curves. After that, period
search and wavelet analysis were applied. We found that:
(i) Fourier analysis found only a median period, but WWZ converged on the
intrinsic rotational period,
(ii) additional spots increased the amplitude and the most significant Fourier
period of the light curve,
(iii) period evolution was gradual if spots were close to each other,
(iv) spots placed almost on the opposite sides of stellar disk generated a rapid
change between $P_{\rm rot}$ and $P_{\rm rot}/2$.

Using WWZ to investigate the period evolution can place further limits on
the possible rotation period. A full model could deal with unknown inclination
of the stellar rotational axis, differential rotation and spot migration.
A light curve with very good sampling and long temporal duration is needed for
such model.

The light curves of HD~283518 (V410~Tau) 
were analyzed sepearately for six individual observing seasons.
The comparison of our study with past data hinted that
the evolution of light curve variations caused by changes in the distribution
of cool photospheric spots may be the result of a $\sim$15-year long cycle
similar to the 11-year cycle of the Sun.

The evolutionary status of our targets was checked using accurate parallaxes and
mean stellar parameters from the literature and calibrations of the  
Str{\"o}mgren-Crawford $uvby\beta$ photometric system. Thanks to the already
published \textit{Gaia} DR2 data, the luminosities are already very accurate
with the largest uncertainty contribution coming from the a-priori unknown
reddening. Furthermore, the uncertainties of the effective temperatures prevent
a suitable calibration of the age and mass to correlate
these parameters with the rotational periods.   

Therefore, we are continuing our investigation on selected targets with dedicated
spectroscopic observations to provide accurate and homogeneous effective temperatures.

\section*{Supporting information}
Additional Supporting Information may be found in the online version of this
article:

\noindent\textbf{Figures A1 and A2:} Fourier plots of investigated targets from ground-based observations

\noindent\textbf{Figure A3:} Fourier plots of targets with available \textit{Kepler} photometry

\noindent\textbf{Figures A4 and A5:} Light curves of investigated targets phase-folded with periods estimated from a single season

\noindent\textbf{Figure A6:} Light curves of targets observed by \textit{Kepler}

\noindent\textbf{Figure A7:} Phased light curves of targets with more than one prominent period in the \textit{Kepler} data

\noindent\textbf{Figure A8:} Evolution of period and amplitude of light curves of our \textit{Kepler} targets

\section*{Acknowledgments}
This study has been supported by the projects VEGA~2/0143/14, VEGA~2/0031/18,
APVV-15-0458 and 7AMB17AT030 (M\v{S}MT). This article was created by the
realisation of the project ITMS No.26220120029, based on the supporting
operational Research and development program financed from the European Regional
Development Fund. This work has made use of data from the European Space Agency (ESA)
mission \textit{Gaia} (\texttt{https://www.cosmos.esa.int/gaia}), processed by
the \textit{Gaia} Data Processing and Analysis Consortium (DPAC,
\texttt{https://www.cosmos.esa.int/web/gaia/dpac/consortium}). Funding
for the DPAC has been provided by national institutions, in particular
the institutions participating in the \textit{Gaia} Multilateral Agreement.
This paper includes data collected by the Kepler mission. Funding for the Kepler
mission is provided by the NASA Science Mission directorate.
Some/all of the data presented in this paper were obtained from the Mikulski
Archive for Space Telescopes (MAST). STScI is operated by the Association of
Universities for Research in Astronomy, Inc., under NASA contract NAS5-26555.
Support for MAST for non-HST data is provided by the NASA Office of Space
Science via grant NNX09AF08G and by other grants and contracts. This research
was made possible through the use of the AAVSO Photometric All-Sky Survey
(APASS), funded by the Robert Martin Ayers Sciences Fund.
We thank the reviewer for his/her thorough review and highly appreciate
the comments and suggestions, which significantly contributed to improving the
quality of the publication.

\bsp

\onecolumn
\newpage
\appendix 
\section{Fourier plots and phased light curves of target stars}

\begin{figure*} 
  \centering
  \includegraphics[width=141mm,clip=0]{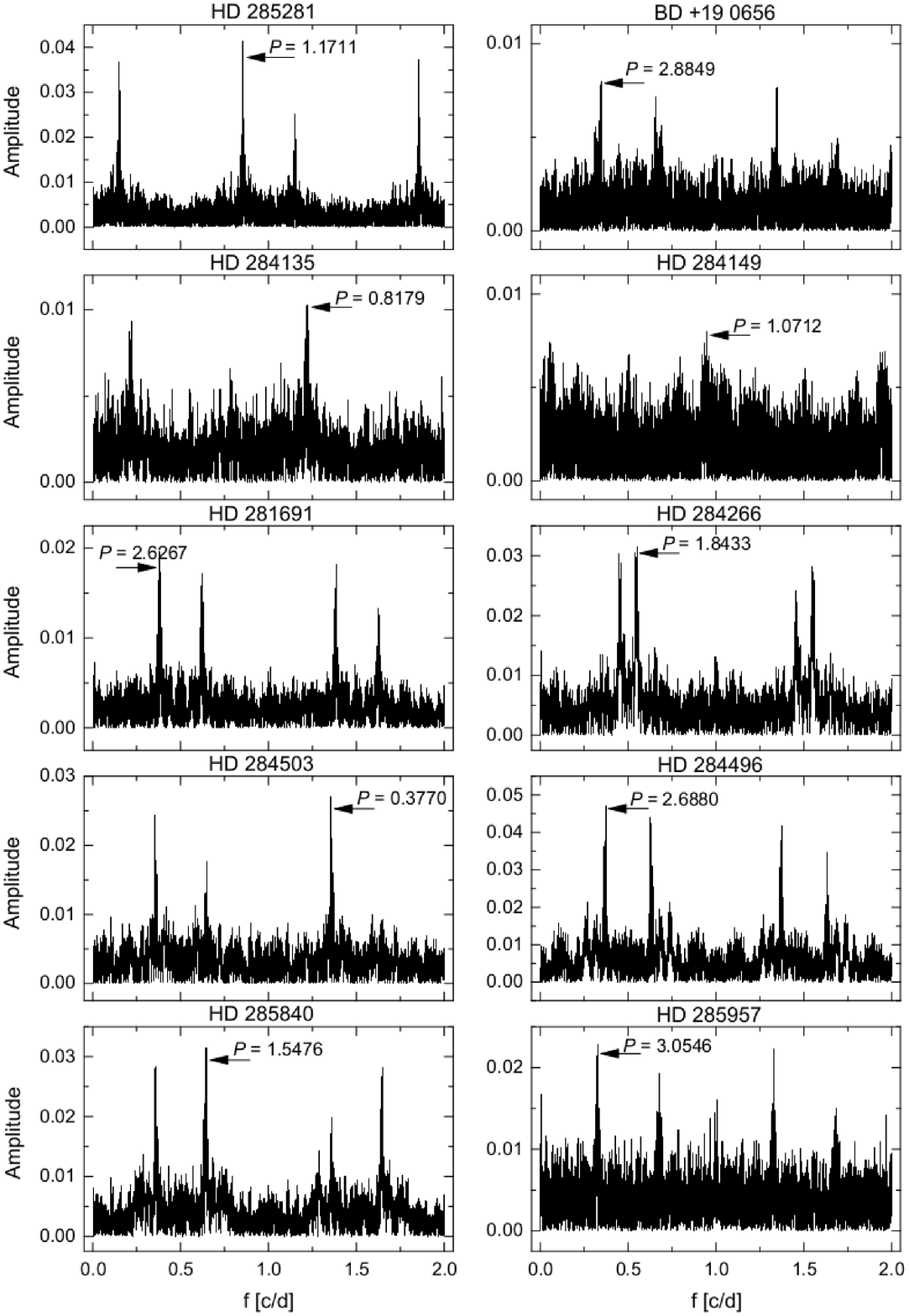} 
  \caption{Fourier plots of investigated targets. Data from all seasons were
  used. Prominent frequencies are noted with an arrow. Details in text.}
  \label{fig:periods1} 
\end{figure*}

\begin{figure*} 
  \centering
  \includegraphics[width=141mm,clip=0]{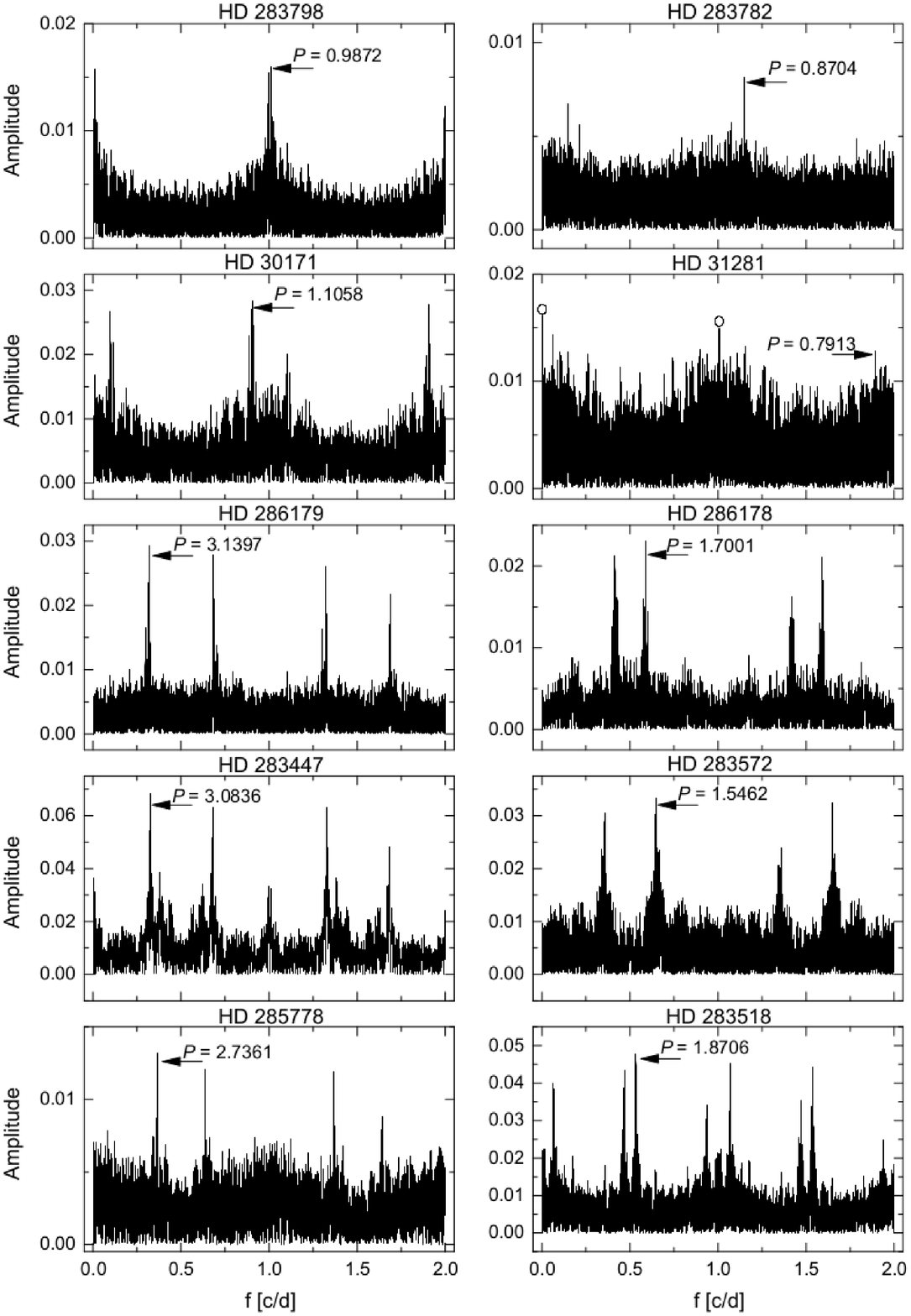} 
  \caption{Fourier plots of investigated targets. Data from all seasons were
  used. Prominent frequencies are noted with an arrow. Details in text.}
  \label{fig:periods2} 
\end{figure*}

\begin{figure*} 
  \centering
  \includegraphics[width=141mm,clip=0]{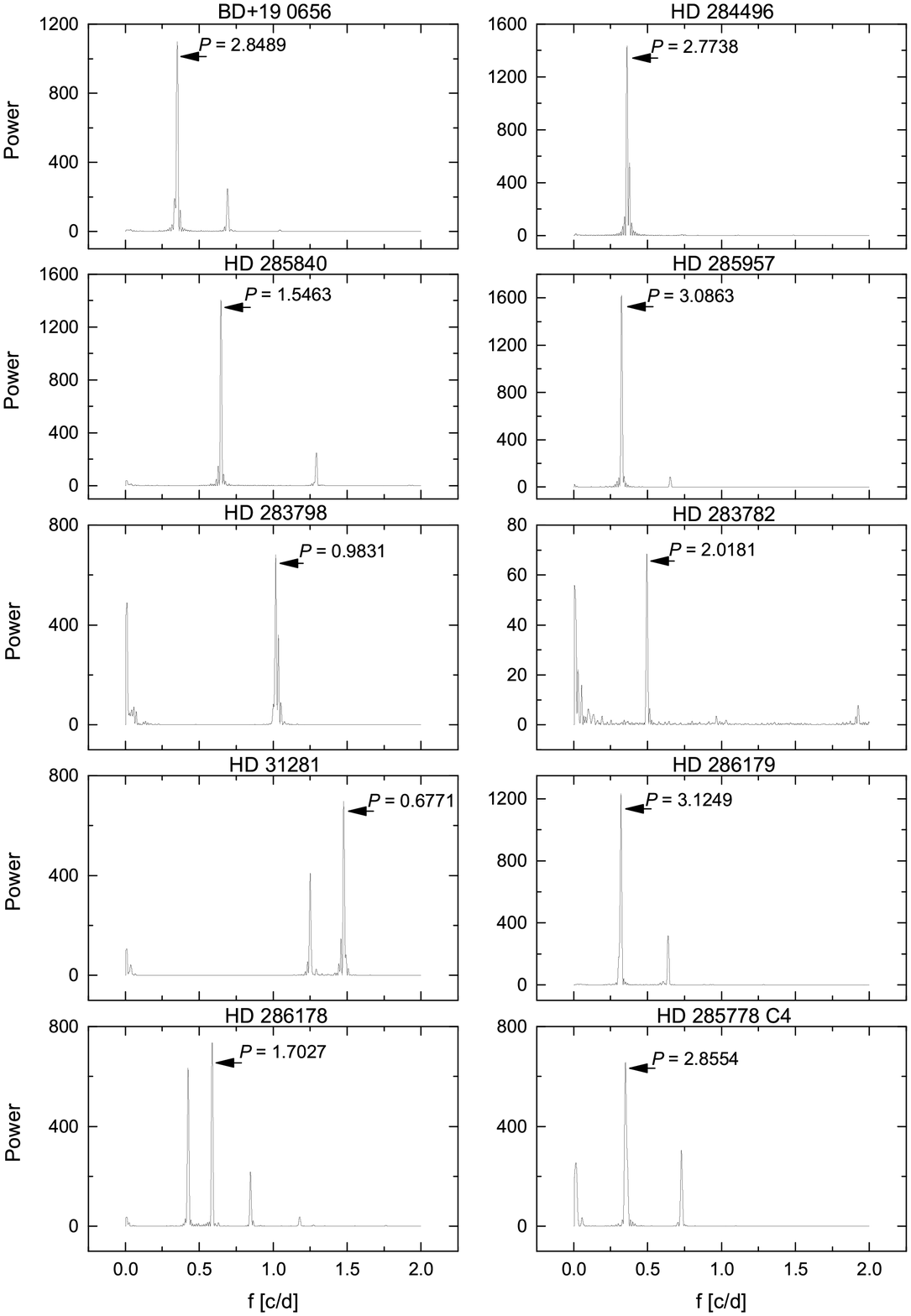} 
  \caption{Fourier plots of targets with available \textit{KEPLER} photometry.
  Prominent frequencies are noted with an arrow. Details in text.}
  \label{fig:periodsK2} 
\end{figure*}

\begin{figure*} 
  \centering
  \includegraphics[width=141mm,clip=0]{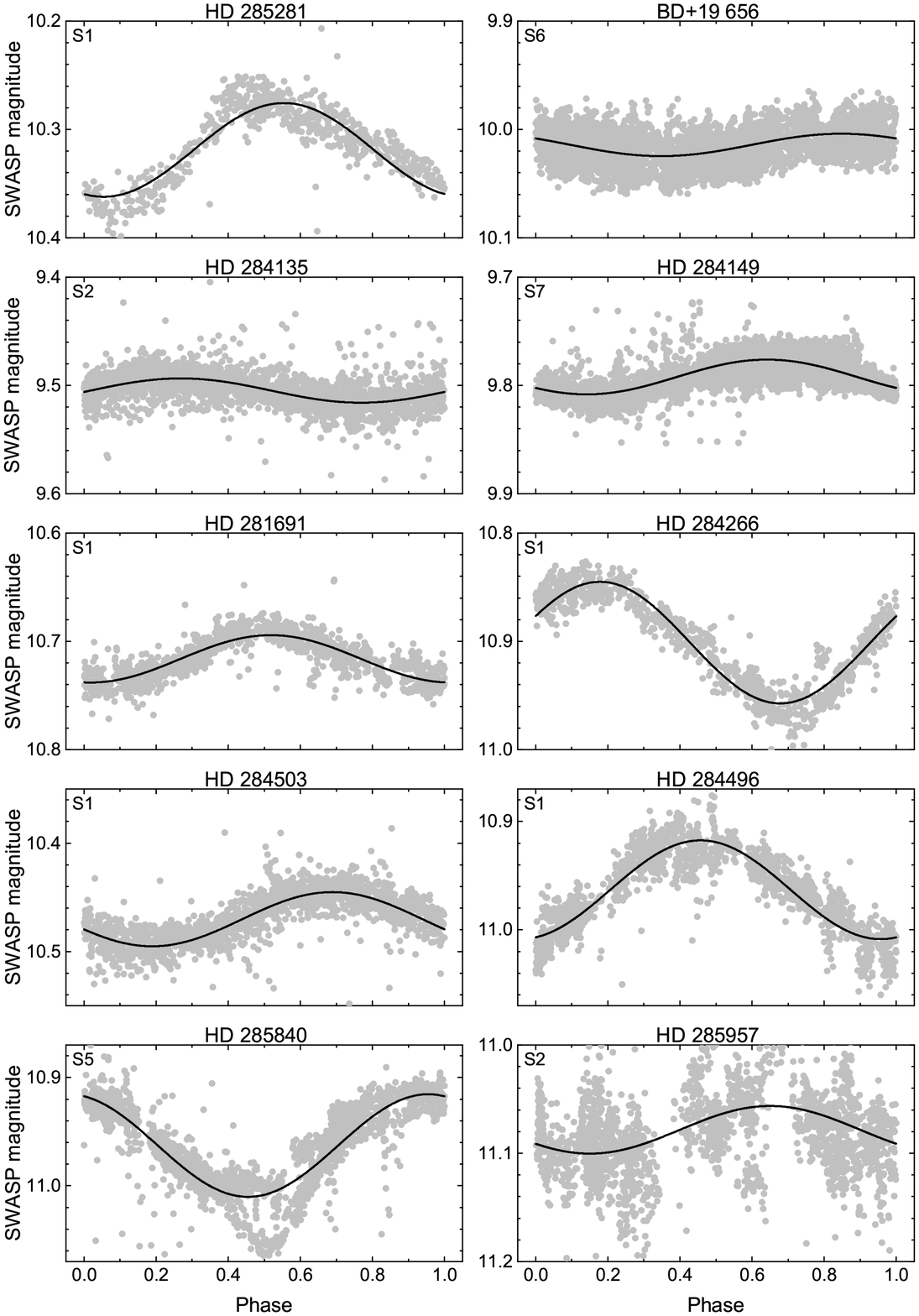} 
  \caption{Light curves of investigated targets phase-folded with periods
  estimated from a single season - indicated by panel number. We have used only
  pre-cleaned data points (see Section \ref{data_swasp}).}
  \label{fig:phases1} 
\end{figure*}

\begin{figure*} 
  \centering
  \includegraphics[width=141mm,clip=0]{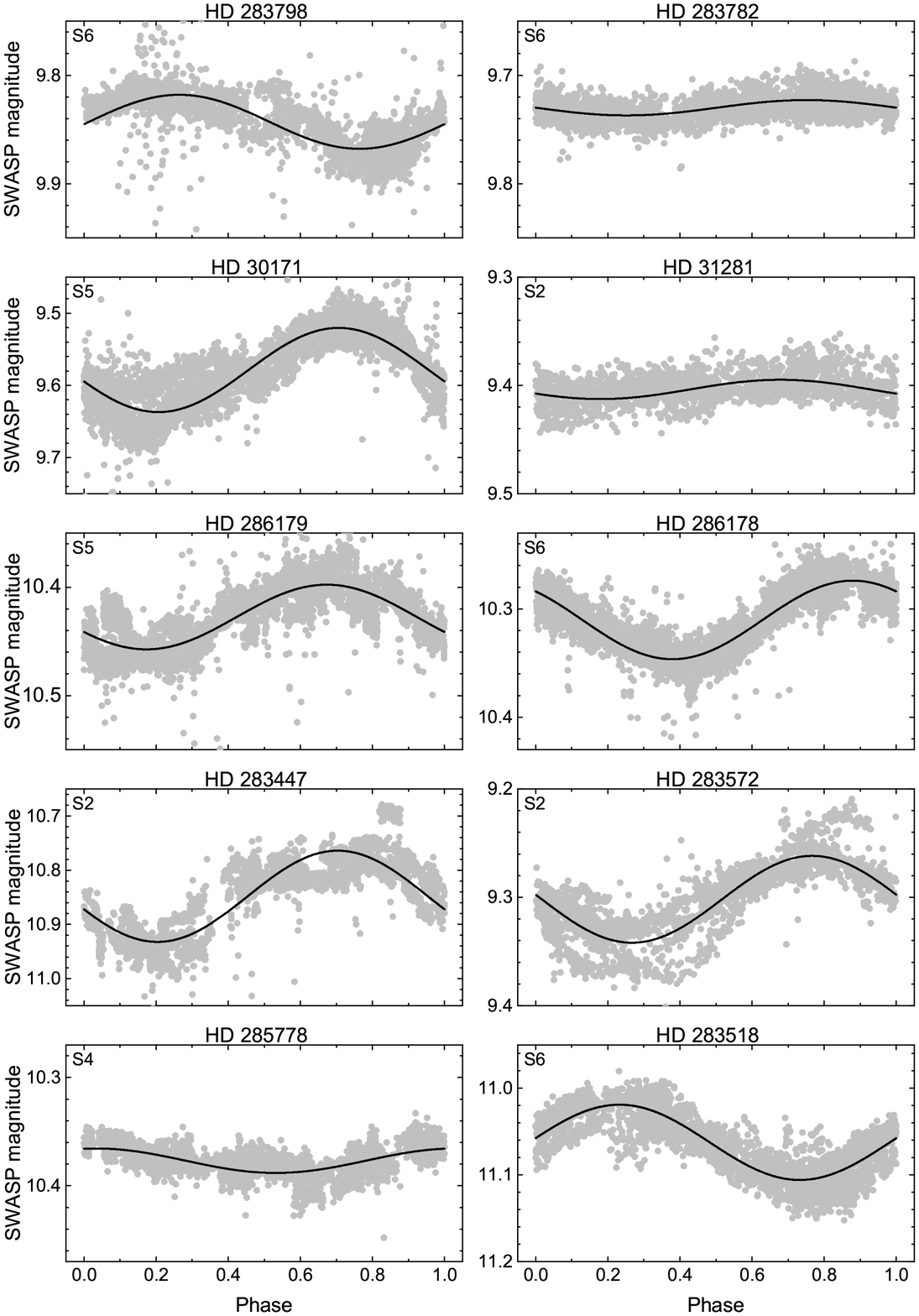} 
  \caption{Light curves of investigated targets phase-folded with periods
  estimated from a single season - indicated by panel number. We have used only
	pre-cleaned data points (see Section \ref{data_swasp}).}
  \label{fig:phases2} 
\end{figure*}

\begin{figure*} 
  \centering
  \includegraphics[width=141mm,clip=0]{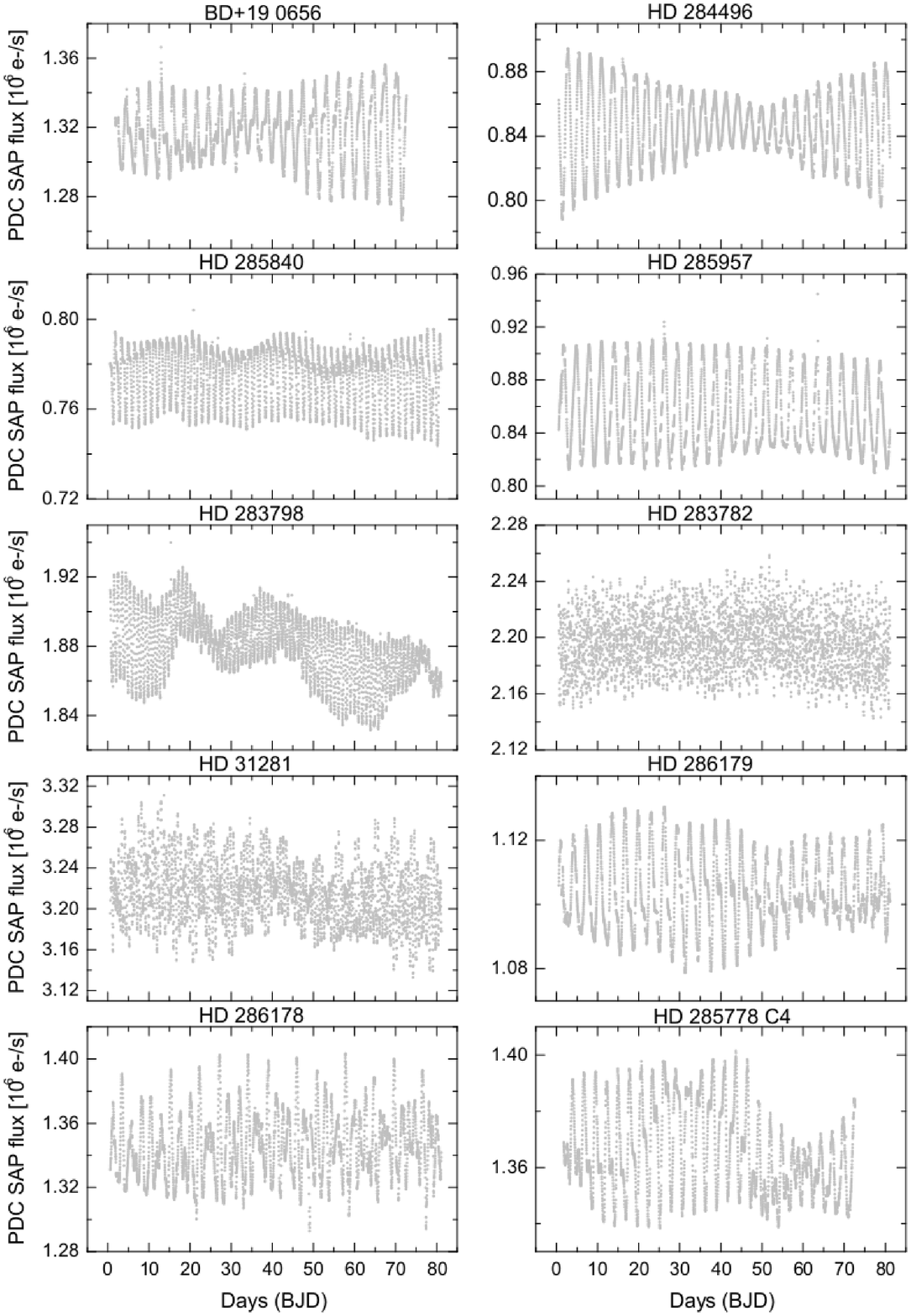} 
  \caption{Light curves of targets observed by \textit{Kepler}.}
  \label{fig:K2LC} 
\end{figure*}

\begin{figure*} 
  \centering
  \includegraphics[width=141mm,clip=0]{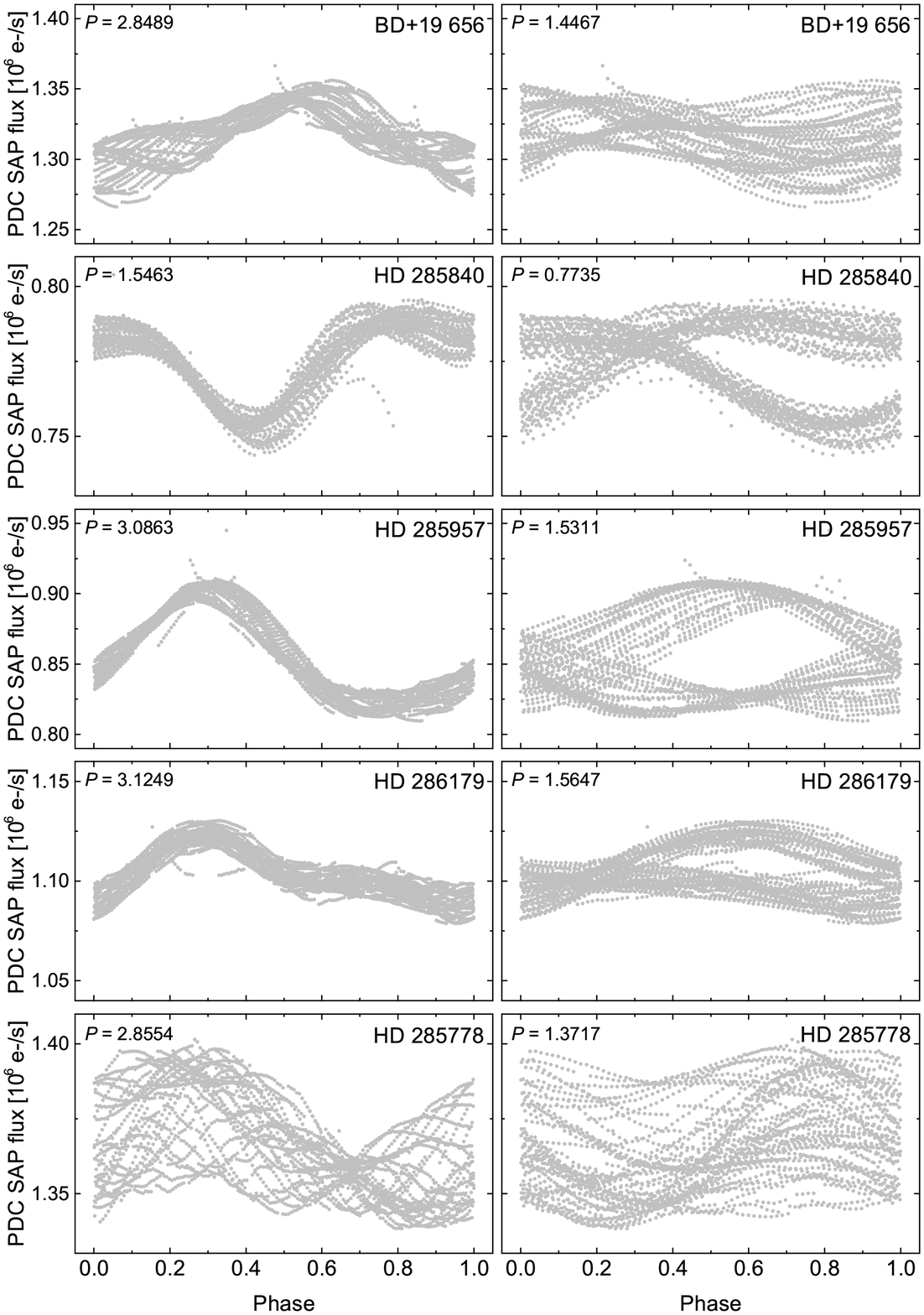}
  \caption{Phased light curves of several stars from our sample for which
  we have found more than one prominent period in the \textit{Kepler} data.
  Data were folded with longer (left) and shorter (right) periods separately.}
  \label{fig:K2phased} 
\end{figure*}

\begin{figure*} 
  \centering
  \includegraphics[width=141mm,clip=0]{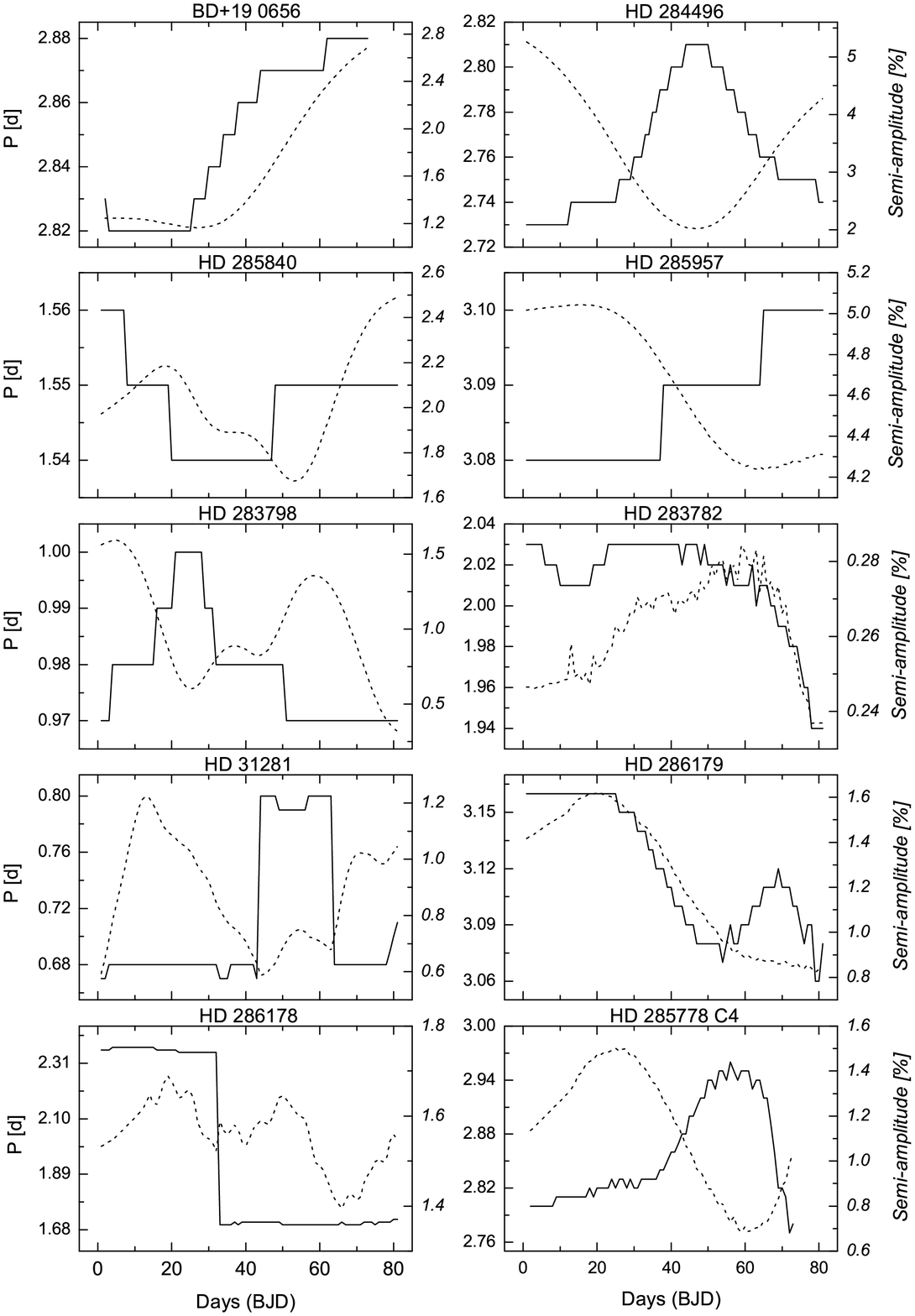} 
  \caption{Evolution of dominant period (solid line) and amplitude (dashed line)
  of light curves of our \textit{Kepler} targets in time. Stepping in periods
  is an artefact of frequency resolution.}
  \label{fig:K2wwz} 
\end{figure*}

\label{lastpage}
\end{document}